\documentclass[11pt, a4paper]{article}

\pdfoutput=1

\usepackage{amsmath}
\usepackage{amsfonts}
\usepackage{amssymb}
\usepackage{cite}
\usepackage{multicol,multirow}
\usepackage{graphicx, physics, subcaption}
\usepackage{graphics, epsfig}
\usepackage{epsf}
\usepackage{epstopdf}
\usepackage[usenames,dvipsnames]{color}

\usepackage{colortbl}
\definecolor{summersky}{cmyk}{0.71,0.33,0,0.14}
\definecolor{flamingo}{cmyk}{0,0.51,0.71,0.14}
\definecolor{rp}{cmyk}{0.2, 1, 0.6, 0}
\definecolor{pacificblue}{cmyk}{0.95,0.3,0, 0.19}
\definecolor{gray60}{cmyk}{0.4,0.4,0,0.8}

\numberwithin{equation}{section}

\usepackage[colorlinks,bookmarks=false, hypertexnames=true]{hyperref}
\hypersetup{
	breaklinks=true,
	pdfstartview={FitH},    
	colorlinks=true,       
	linkcolor=blue,          
	citecolor=red,        
	filecolor=magenta,      
	urlcolor=red,           
	anchorcolor=green,      
	linktocpage=true
}

\newcommand{\nc}{\newcommand}

\nc{\ba}{\begin{eqnarray}}
\nc{\ea}{\end{eqnarray}}

\nc{\calR}{{\cal{R}}}
\nc{\calP}{{\cal{P}}}
\nc{\cN}{ {\cal{N}} }

\begin{document}

\def\thefootnote{\fnsymbol{footnote}}

\begin{center}

{\bf  Bound State Perturbations in the Interior of Black Holes 
}
\\[0.5cm]

{
Hassan Firouzjahi\footnote{firouz@ipm.ir}
}
\\[0.5cm]
 
 {\small \textit{School of Astronomy, Institute for Research in Fundamental Sciences (IPM) \\ P.~O.~Box 19395-5746,, Tehran, Iran
}}\\

\end{center}

\vspace{1cm}
\hrule
\begin{abstract}
We revisit our earlier work and investigate the bound state perturbations in the interior of the Schwarzschild  black hole. The bound sates are defined as the perturbations in the interior of the black hole 
with an imaginary spectrum which are regular at the center of black hole while their time-dependent profile falls off exponentially on the event horizon. Using the scale factor in the expanding direction in the interior of the black hole as the clock, we rewrite the corresponding Regge-Wheeler equation and solve it semi-analytically as well as numerically. We confirm that the  bound state solutions exist for scalar, vector and axial tensor perturbations. It is shown that for a given value of $\ell >s$, there are total $\ell-s$ such bound states.  We obtain the universal lower bound $2 G M \omega_I >1$ for the spectrum of bound state  which is asymptotically saturated in the large $\ell$ limit.  Furthermore, we obtain an upper bound on the spectrum 
of axial perturbations which for large $\ell$ scales like $2 G M \omega_I \lesssim 0.04\,  \ell^4 $.  
As  observed recently, these bound states have the curious property that the profile of the total wave function has a non-zero 
magnitude near the future event horizon.

\end{abstract}

\vspace{0.3cm}
\hrule

\newpage

\section{Introduction}
\label{sec:intro}

The black hole (BH) physics is a central theme of the modern physics literature both theoretically and observationally. On the observational side,  
the detections of gravitational waves 
by LIGO, Virgo and KAGRA  from the merging of BHs of different masses have put the reality of BHs in cosmos beyond doubt 
\cite{LIGOScientific:2016aoc, LIGOScientific:2017vwq, LIGOScientific:2020zkf, LIGOScientific:2025obp}. On the theoretical side, understanding the mechanism of the formations  of BHs and supermassive BHs (wether astrophysical or primordial) along with the nature of singularity at the center of BH and its link to quantum gravity and questions such as Hawking radiation, information paradox etc  are active areas of research. 

A central role in the studies of the BHs is played by quasi normal modes (QNMs) which describe the response of a BH to the exterior perturbations in its final stages, for a review see  \cite{Nollert:1999ji, Kokkotas:1999bd, Berti:2009kk, Berti:2025hly} and the references therein. The QNMs are characterized by specific boundary conditions for the perturbations in the exterior region of BH. More specifically, the perturbations are assumed to be purely outgoing on far spatial distances and ingoing on the surface of the event horizon. These boundary conditions fix the spectrum of perturbations. The perturbations of massless scalar, vector and axial tensor perturbations are governed by the Regge-Wheeler (RW) equation  while the polar  perturbations are described by the Zerilli equation \cite{Regge:1957td, Zerilli:1970se, Chandrasekhar:1985kt}. The solutions of the RW and Zerilli equations can not be expressed in terms of basic special functions so the spectrum are obtained mostly numerically \cite{Chandrasekhar:1975zza, Leaver:1985ax, Anderson:1991kx, Nollert:1993zz}. For an incomplete list of 
various theoretical studies to calculate the spectrum of the QNMs for the Schwarzschild BH see \cite{Mashhoon:1982im, Blome:1981azp, Ferrari:1984zz, Ferrari:1984ozr, Liu:1996cxr, Schutz:1985km, Iyer:1986np, Iyer:1986nq, Motl:2003cd, Andersson:2003fh, Berti:2005ys, Hatsuda:2019eoj, 
Volkel:2025lhe, Hod:2025scz}.

The perturbations in the exterior regions are vastly studied, owing to the fact that these perturbations are  directly related to observational detections such as the gravitational waves from the merging of BHs. Having said this,  significantly less studies are devoted for the perturbations in the 
interior of BHs.   The prime reason is that the interior region is causally disconnected to the exterior region. A unique property of BH is that the central singularity is separated from the exterior region by the surface of the event horizon such that the interior region is casually disconnected from the exterior region. Any physical effect occurring in the interior region is inaccessible to the exterior observers. However, from the theoretical point of view,  it is quite natural to examine what happens for the perturbations which are confined to the interior region. This question was originally studied in  \cite{Firouzjahi:2018drr} who studied the bound state perturbations in the interior of the Schwarzschild BH. The bound states are defined as the perturbations with an imaginary spectrum 
which are regular in the center of BH while falls off exponentially on the surface of the event horizon. Furthermore, the possible links between the spectrum of bound state perturbations and the spectrum of QNMs associated for the exterior perturbations are studied in \cite{Firouzjahi:2018drr}. The question of bound state perturbations  in the interior of  the Schwarzschild  BH was studied more recently in \cite{Steinhauer:2025bbs} as well. For other works concerning the perturbations in the interior of BHs see \cite{Fiziev:2006tx, Firouzjahi:2024xbo}. 

Another theoretical motivation to study the perturbations in the interior of BH is that the interior of BH is like an anisotropic cosmological setup. More specifically, the interior of the Schwarzschild BH is like the Bianchi I universe in which the azimuthal directions are collapsing  but the extended spatial direction is expanding as one approaches the center of BH.  As such, the study of the perturbations in this  anisotropic  cosmological setup is rich and non-trivial. One may borrow insights and techniques from cosmological perturbations theory in  FLRW cosmology which is directly related to primordial fluctuations from inflation and structure  formation in cosmos  \cite{Kodama:1984ziu, Mukhanov:1990me}.

In this work, we revisit our earlier work \cite{Firouzjahi:2018drr} and study the bound state perturbations in the interior of Schwarzschild BH for scalar, vector and axial tensor perturbations. We calculate the spectrum of these bound states solutions   semi-analytically as well as numerically and discuss their physical implications.  We comment that the bound state perturbations in the interior of BHs considered here  are different than the bound state of the inverted potential pioneered by Mashhoon and collaborators  \cite{Mashhoon:1982im, Blome:1981azp, Ferrari:1984zz, Ferrari:1984ozr, Volkel:2025lhe}. The bound states in \cite{Mashhoon:1982im, Blome:1981azp, Ferrari:1984zz, Ferrari:1984ozr} are based on  the complex transformation of the RW equation to bring the potential in a form such as the Poschl-Teller or the Eckart potentials which admit analytic solutions. 
This approach was further pursued recently in \cite{Volkel:2025lhe} who studied the inverted  RW potential directly. See also \cite{Hatsuda:2019eoj}
for the study of bound states concerning the inverted potential of the RW equation.  In contrast, in our study,  the bound state is interpreted as the solution with an imaginary frequency to the RW equation in the interior of BH in which its profile
is confined to the interior of the BH.

The rest of the paper is organized as follows. In section \ref{interior}
we review the geometry of the interior of the Schwarzschild BH as an anisotropic cosmological background. In section \ref{perts} we study the perturbations and the corresponding RW equation in various coordinate systems and look for the shapes of the potential which can support the bound states. In sections \ref{WKB} and \ref{matching} we present our approximate analytical methods to calculate the spectrum analytically  while the  numerical method to calculate the spectrum is presented  in section \ref{numerical}. In section \ref{implications} we review the physical implications of the bound states and compare our results with those of more recent work \cite{Steinhauer:2025bbs}, followed by the Summary and Discussions in section \ref{summary}.

\section{Geometry of the Interior of Black Holes}
\label{interior}

Let us start with  the Schwarzschild metric in its usual spherical form,  
\ba
\label{metric-Sch}
\dd s^2 = - \big(
1- \frac{2GM}{r}\big) \dd { t\, }^2 + \frac{\dd r^2}{\big(
	1- \frac{2GM}{r}\big)} + r^2 \dd \Omega^2 \, ,
\ea
in which $G$ is the Newton constant, $M$ is the mass of the BH  and  $\dd\Omega^2$ represents the angular parts of the metric. As it is 
well-known,  the coordinate system $(r, t)$  suffers from a coordinate singularity  at the event horizon $r= 2 GM$. Furthermore, in the interior of the BH, the roles of $(r, t)$ coordinates change in which $ t$ becomes spacelike while $r$ becomes timelike. To avoid the coordinate singularity 
on the event horizon,  one can use the  Kruskal-Szekeres coordinate, such as its lightcone variant $(U, V)$ coordinate, which covers the entire manifold. 

As the roles of $(r, t)$ coordinates are switched in the interior region, the dynamics of the interior of the BH with the metric (\ref{metric-Sch}) represents a cosmological setup. The angular directions are shrinking as one approach the singularity while the spatial direction along the $t$ coordinate is expending. This suggests that the interior of BH is an anisotropic cosmological background like the Bianchi type I metric, known as the Kantowski-Sachs background \cite{Kantowski:1966te, Doran:2006dq}.  Furthermore, the singularity at $r=0$ represents a big crunch, a spacelike singularity in the future.

To study the dynamics of interior, let us 
start with the usual  tortoise coordinate, 
\ba
\label{r*r-eq}
\dd r_* = (1- \frac{2GM}{r})^{-1} \dd r \, .
\ea
Note that the above differential relation is valid for both interior $r< 2 G M$ and the exterior $r > 2 GM$ regions. For the  interior of BH, 
the above definition yields, 
\ba
\label{r*-BH}
r_* = r + 2 GM \ln \left( 1- \frac{r}{2GM} \right) ,   \qquad \quad ({r < 2 GM}) \, .
\ea
The  horizon  $r=2GM $ is mapped to  $r_*=-\infty$ 
while the singularity at $r=0$ corresponds to $r_* =0$.

Related to $\dd r_*$, we can define  the future directed time coordinate $\dd\tau=\dd r_*$ so Eq. (\ref{r*-BH}) can be written as, 
\ba
\label{tau-def}
\tau = r(\tau) + 2 GM \ln \Big( 1- \frac{r(\tau)}{2GM} \Big) \qquad \quad ({r < 2 GM}) \, .
\ea
Note that from the above equation, one can solve for $r$ as a function 
of $\tau$.

With this definition,  the BH singularity is located at $\tau=0$ while $\tau=-\infty$ represents the position of  the event horizon. As the $t$ coordinate is spacelike in the interior, let us for the convenience in our analysis,  define $\dd x \equiv \dd t$ with $-\infty < x < +\infty$. Correspondingly,  the metric of the interior of BH in terms of $(\tau, x)$ coordinates  takes the following cosmological form, 
\ba
\label{BH-metric}
\dd s^2 = a(\tau)^2 ( -\dd \tau^2 + \dd x^2) + r(\tau)^2 \dd \Omega^2 \, ,
\ea
where the scale factor $a(\tau)$ is given by, 
\ba
\label{a-scale}
a(\tau) \equiv  \Big( \frac{2 G M}{r(\tau)} -1  \Big)^{\frac{1}{2}} \, . 
\ea
The metric 
(\ref{BH-metric}) represents an anisotropic cosmological background which has  two scale factors $a(\tau)$ and $r (\tau)$ which have different dynamics.   Near the horizon region $\tau \rightarrow -\infty$,  we have $a(\tau) \rightarrow 0$ while $r(\tau) \rightarrow 2 G M$. On the other hand, as time proceeds, the 
scale factor along the two-sphere with the scale factor $r(\tau)$ shrinks while 
the space along the $x$ direction  with the scale factor $a(\tau)$ expands.  
Finally, at the time of big crunch when $\tau=0$,  $r(\tau) \rightarrow 0$ while  $a(\tau) \rightarrow \infty$. For further aspects of the cosmology of the interior of BH see \cite{Firouzjahi:2024xbo, Firouzjahi:2022rtn}.

It is convenient to rescale the coordinates via $x\equiv 2 GM \bar x, r \equiv 2 GM \bar r, \tau \equiv 2 GM \bar \tau$ so $\dd s^2= (2 GM)^2 {\bar{\dd s}}^2 $ while from now on we drop the overline symbol for the convenience (unless mentioned otherwise).   With this convention, the horizon is located at 
$r=1$. 

As in FLRW cosmology, it is very helpful  to use the 
scale factor $0< a(\tau) < \infty$ as the clock. This is viable since $a(\tau)$ is a monotonically increasing function in the interior of the BH and there is a one-to-one relation between $\tau$ and $a$. More specifically, from Eq. (\ref{a-scale}) we obtain,
\ba
\label{r-a}
r= \frac{1}{1+ a^2} \, ,
\ea
while from the definition of the tortoise coordinate (\ref{tau-def}), we have,
\ba
\label{tau-a}
\dd \tau= -\frac{\dd r}{a^2} = -\frac{2 \dd a}{a (1+ a^2)^2} \, .
\ea
Plugging the above relations into metric (\ref{BH-metric}), yields, 
\ba
\label{BH-metric-a}
\dd s^2 = -\frac{4 \dd a^2}{(1+ a^2)^4} + a^2 \dd x^2 + \frac{1}{ (1+ a^2)^2} \, \dd \Omega^2 \, .
\ea
The above metric should be viewed in terms of $a$ as an independent variable. In other words, our coordinates now are $(a, x, \theta, \phi)$, in which  $a$ runs in the interval $0< a < \infty$ while  $-\infty < x < \infty$ as before.  One advantage of the metric (\ref{BH-metric-a}) with coordinate $a$ is that its components  are algebraic while the relation between $\tau$ (or  $r_*$) and $r$ in the original coordinate (\ref{BH-metric}) involves a transcendental equation.   

We can use the metric (\ref{BH-metric-a}) to solve for the spectrum of the bound states.  However, as in FLRW cosmology, it is convenient to use the number of e-folds $N$ associated to $a$ as the clock.  
The number of e-fold is defined via  $a(N)\equiv e^{N}$  with $ -\infty < N <\infty $  in which the event horizon is mapped to $N=-\infty$ while the singularity at $r=0$ corresponds to $N=+\infty$. In terms of the number 
of e-folds $N$, the metric (\ref{BH-metric-a}) is cast into, 
\ba
\label{BH-metric-N}
\dd s^2 = -\frac{4 e^{2 N} \dd N^2}{(1+ e^{2 N})^4} + e^{2 N} \dd x^2 + \frac{1}{ (1+ e^{2 N})^2}\,  \dd \Omega^2 \, . 
\ea

Alternatively, we can use the angular parametrization of the scale factor via,
\ba
\label{a-chi}
a(\chi)= \tan(\chi)  \rightarrow r= \cos(\chi)^2 \, ,  
\ea
in which  the metric takes the following form,
\ba
\label{BH-metric-chi}
\dd s^2 = \cos^4(\chi) \big( - 4 \dd \chi^2 + \dd \Omega^2)  + \tan^2(\chi) \dd x^2\, .
\ea 
One can get rid of the factor $4$ above via rescaling of $\chi$ and make the metric along the three directions $(\chi, \theta, \phi)$ conformal, but we prefer to keep it as above in order to keep the periodicity of $\chi$ simple. 
As we see, the metric in terms of the $\chi$ coordinate has a reasonably simple form. The position of the horizon  $r=1$ with $ a=0$ corresponds to $\tan(\chi)=0$. The independent solutions are $\chi= k \pi$ with $k=0, 1$, assuming that  $\chi \cong \chi+ 2 \pi$. Demanding that $a(\chi) >0$ requires
$0< \chi < \frac{\pi}{2}$. Correspondingly, the horizons (past and future horizons) are mapped to $\chi=0$ while the singularity at the center of BH is mapped to $\chi=\frac{\pi}{2}$. On the other hand, the periodic structure of
$\tan(\chi)$ suggests that the coordinate $\chi$ maybe extended to other parts of the manifold such as the white hole region as well. It would be interesting to look for this maximal extension in terms of the $\chi$ coordinate.


\section{Perturbations in the Interior of Black Holes}
\label{perts}

Here we study the perturbations in the interior of the BHs. We consider the scalar, vector and axial tensor perturbations with spin $s=0, 1,2$ respectively. The master equation  associated to  the evolution of these perturbations is given by the Regge-Wheeler (RW) equation 
\cite{Regge:1957td}. The equation for the polar tensor perturbation is governed by the Zerilli \cite{Zerilli:1970se} equation which is not considered here. 

\subsection{Regge-Wheeler Equation and Effective Potential }

Traditionally, the RW equation is written for the perturbation outside the BH with $r>1$. However, formally, the RW equation is valid in the interior of the BH as well, only one has to be careful with its interpretation as the role of $(r,t)$ coordinates in the original Schwarzschild metric are switched. Specifically, suppose we expand the perturbations in terms of the spherical harmonics $Y_{\ell m}$ as usual with $\ell$ representing  the index of spherical harmonics. Since the spatial direction $x$ enjoys  translation invariance we can decompose the remaining parts of the perturbation in Fourier space as $Z(x, \tau) \propto  e^{-i \omega x} Z (\tau)$. Then, the 
RW equation in $\tau$ coordinate is written as, 
\ba
\label{eq-pert}
\frac{\dd^2}{\dd \tau^2}  Z(\tau) + \big( \omega^2 - V_{\mathrm{eff}}(\tau) \big) Z(\tau) = 0 \, ,
\ea 
with the effective potential given by \cite{Firouzjahi:2024xbo, Firouzjahi:2022rtn}, 
\ba
\label{Veff_spin}
V_{\mathrm{eff}} (\tau) = \Big(1- \frac{1}{r(\tau)} \Big)  \left[ \frac{\ell (\ell+1)}{r(\tau)^2} + \frac{(1-s^2)}{r(\tau)^3} \right] \, .
\ea 
We have rescaled the frequency in unit of $2 GM$, i.e.  
$\omega \rightarrow 2 G M \omega$,  as the coordinates are already rescaled by $2 GM$.  Here we have started with the conformal time $\tau$ but later on we switch to other appropriate clocks such as $N$ and $\chi$ when required. 

As $\tau$ is timelike, the RW equation (\ref{eq-pert}) represents the dynamical evolution of the perturbation inside the BH. In this view, its interpretation   is fundamentally different than the usual RW equation for the perturbations propagating in the 
exterior region. In a sense, Eq. (\ref{eq-pert})  may be viewed as the extension of the Mukhanov-Sasaki equation in cosmological perturbation theory, now written in an anisotropic cosmological background. The quantum aspects of the perturbations inside the BH and inside the white hole were studied in \cite{Firouzjahi:2024xbo} and 
\cite{ Firouzjahi:2022rtn} respectively.  Here, we only consider Eq. (\ref{eq-pert}) at the classical level, similar to standard QNMs analysis.


\begin{figure}[t]
\begin{center}
	\includegraphics[scale=0.32]{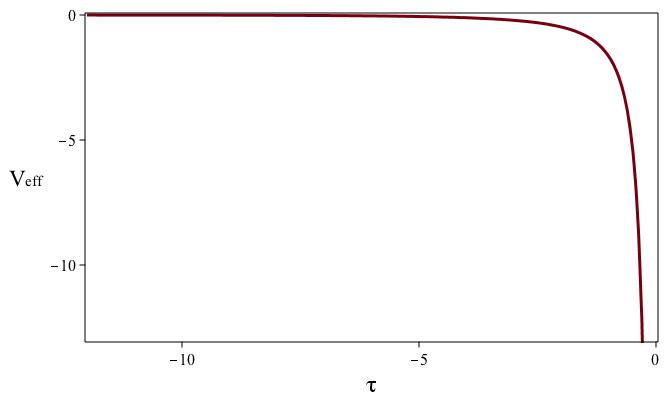}
	\hspace{1cm}
	\includegraphics[scale=0.32]{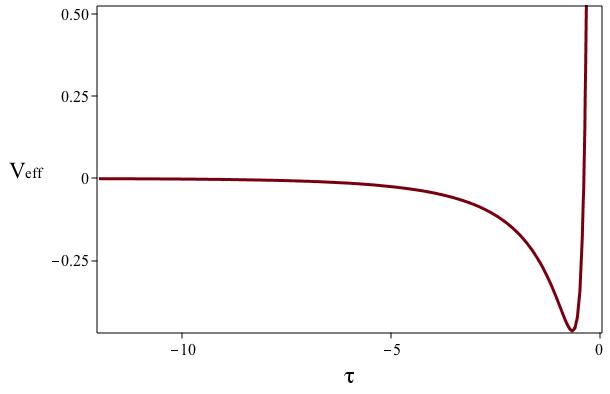}
	\end{center}
\caption{ The effective potential $V_\text{eff}(\tau)$ given in Eq. (\ref{Veff_spin}) for $s=0$ (Left) and $s=2$ (Right) with $\ell=2$ in both plots. 
\label{Vef}
}
\end{figure}

The shape of $V_\text{eff}(\tau)$ for $s=0$ and $s=2$ are plotted in Fig. \ref{Vef}. As $\tau \sim -r^2/2$ near the center, one concludes that the potential scales like $(s^2-1)/4\tau^2$ at the center. 
For $s=0$ the potential is negative and monotonically decreasing while for $s=2$  the potential develops a global minimum and approaches $+\infty$ near the center. 
Since   $V_\text{eff}(\tau)$ has a negative global minimum for axial perturbations, the bound state solutions with $\omega^2 <0$ can be supported  for these perturbations. On the other hand, the existence of  bound  states for scalar perturbations with a negative  potential as in the left panel of Fig. \ref{Vef} may not be obvious. However, the situation here is 
qualitatively like the s-wave states of Hydrogen atom and the bound states with $\omega^2 <0$  can exist.  See \cite{Gangopadhyaya:2017wpf} for further discussions for bound state solutions in singular potentials. 
Using some approximation expression for the effective potential $V_{\mathrm{eff}} (\tau) $, the spectrum of $\omega$ for bound state solutions are studied in \cite{Firouzjahi:2018drr}. It was advocated in \cite{Firouzjahi:2018drr} that the spectra of bound state perturbations may be related to   the spectra of QNMs.

As the relation between $r$ and $\tau$ is transcendental, it is not possible to write a simple algebraic form for  $V_{\mathrm{eff}} (\tau)$ in terms of $\tau$.  To bypass this difficulty, we employ other variables as the clocks such that the corresponding effective potentials  have algebraic forms. 
More specifically, we use the scale factor $a$ as the clock, and use either the number of e-fold $N$ or its angular representation $\chi$ to rewrite the RW equation which yield algebraic expressions for the corresponding effective potentials. 

As we change the coordinate from $\tau$ to $a$, a ``friction" term like $\partial_a Z(a)$  will appear. To get rid of this first derivative term, let us define the auxiliary field $A$ via,
\ba
Z(a)\equiv \Big[(a^2+ 1)\sqrt{a} \Big]^{-1} A(a) \, .
\ea
Then the RW equation in terms of the clock $a$ is written as,
\ba
\label{RW-a}
\frac{\dd^2 A(a)}{\dd a^2} - {U_\text{eff}}(a) A(a)=0 \, ,
\ea  
in which the new effective potential $U_\text{eff}$ is given by,
\ba
\label{Ueff-a}
U_\text{eff}(a) \equiv -\frac{4 \omega^2}{a^2 (a^2+ 1)^4} -  \frac{4 \ell (\ell+1)}{(a^2+ 1)^2} -\frac{(1-16 s^2 ) a^2 +1}{4 a^2 (a^2+ 1)} \, .
\ea
As expected, the effective potential given above has the algebraic form.  However, we notice that the frequency $\omega^2$ is now absorbed into the effective potential. This means that one can interpret Eq. (\ref{RW-a}) as a Schrodinger-like equation with the ``energy" equal to zero. Looking at the structure of $U_\text{eff}(a)$, we notice that in the absence of $\omega$, the solution can be written in terms of the hypergeometric functions. However, with the effect of $\omega$ included, this property is lost and the solutions are given  in terms of the Heun functions which are not particularly helpful when we impose our initial  and boundary conditions. 

It is also useful to look at the asymptotic values of the effective potential. Near the horizon, $a\rightarrow 0$, it blows up as,
\ba
\label{ap-a1}
U_\text{eff}(a) \simeq -\big( 4 \omega^2 +\frac{1}{4} \big) a^{-2} \, ,
\quad \quad (a \rightarrow 0) \, ,
\ea
while near the singularity, $a \rightarrow \infty$, it approaches zero as,
\ba
\label{ap-a2}
U_\text{eff}(a) \simeq
\big( 4 s^2 -\frac{1}{4} \big) a^{-2} \, ,
\quad \quad (a \rightarrow \infty) \, . 
\ea
An important conclusion can be drawn here. In order for the bound state 
solution of the Schrodinger-like equation (\ref{RW-a}) with the effective zero energy to exist, we require that near the horizon ($a\rightarrow 0$), 
the effective potential $U_\text{eff}(a)$  to be positive.  Assuming that $\omega$
is pure imaginary with $\omega= i \omega_I$,  this immediately yields the lower bound  $|\omega_I |> \frac{1}{4} $. We will confirm this lower bound with other methods as well.

Alternatively, we can use the number of e-folds $a= e^N$ or the angular representation $a= \tan(\chi)$ as the clock. Using  the $N$ coordinate as the clock  and defining the new field $A(N)$ as,
\ba
\label{Z-A}
Z(N)\equiv \sqrt{\frac{a'}{a}} (1+ a^2)^{-1} A(N)\, ,
\ea
with prime indicating the derivative with respect to $N$, the corresponding Schrodinger-like equation is given by,
\ba
\label{RW-N}
\frac{\dd^2 A(N)}{\dd N^2} - {U_\text{eff}}(N) A(N)=0 \, ,
\ea  
where now the effective potential $U_\text{eff}(N)$ is given by,
\ba
\label{Ueff-N}
U_\text{eff}(N) =  -\Big[ \frac{4 \omega^2}{(1+ e^{2 N})^4} + 
\frac{4 \ell (\ell+1)  e^{2 N}}{(1+ e^{2 N})^2} -  
\frac{4 s^2  e^{2 N}}{1+ e^{2 N}} \Big] \, .
\ea
Looking at the asymptotic regions, near the horizon $N\rightarrow -\infty$, 
and near the singularity $N \rightarrow \infty$, it reaches two constant values,
\ba
\label{ap-N}
U_\text{eff}(N\rightarrow -\infty) \simeq -4 \omega^2  \, , 
\quad \quad
U_\text{eff}(N\rightarrow +\infty) \simeq  4 s^2 \, .
\ea

\begin{figure}[t]
\begin{center}
	\includegraphics[scale=0.21]{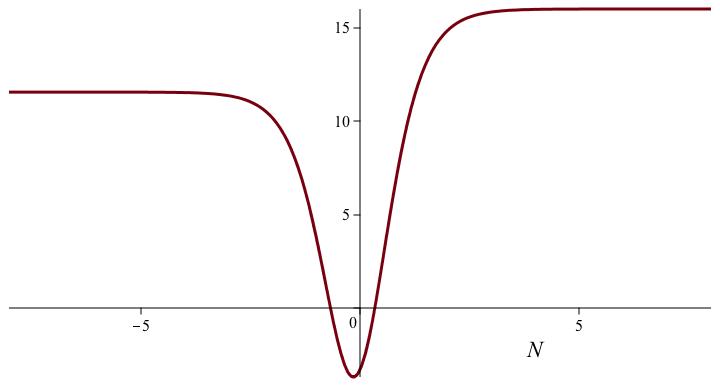}
	\hspace{0.3cm}
	\includegraphics[scale=0.21]{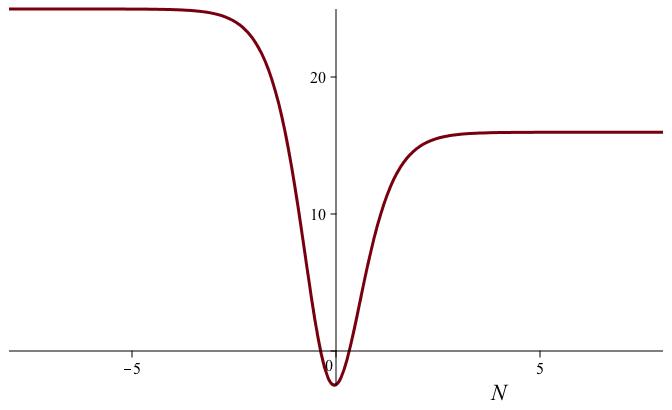}
	\hspace{0.3cm}
	\includegraphics[scale=0.21]{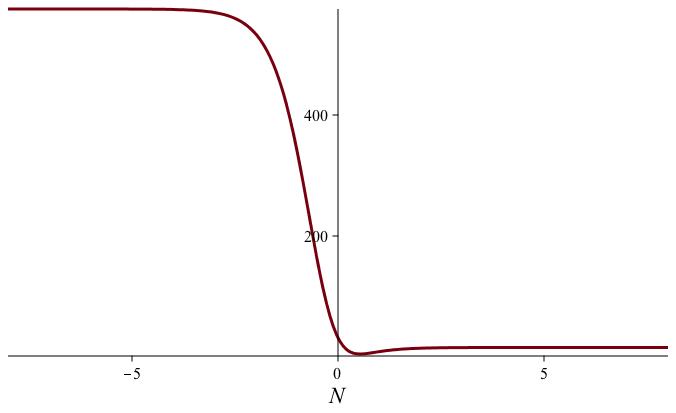}
	\end{center}
\caption{ The effective potential $U_\text{eff}(N)$ given in Eq. (\ref{Ueff-N})
 for $s=2$ and $\ell=3$ with $\omega_I= 1.7$ (Left), $\omega_I= 2.5$ (Middle) and $\omega_I= 12 $ (Right).  As $\omega_I$ increases the global minimum is uplifted so for large $\omega_I$ the bound state does not exist. 
 The asymptotic behaviours are given by Eq. (\ref{ap-N}). 
\label{Ueff-N-plot}
}
\vspace{0.5cm}
\end{figure}

For a view of $U_\text{eff}(N)$ for the axial tensor perturbations ($s=2$) with $\ell=3$  see Fig. \ref{Ueff-N-plot}.
Note that in these plots, $\omega^2$ is part of the effective potential while in the corresponding Schrodinger equation the effective energy is zero. The shape of the potential depends on $\omega$.  For small $|\omega_I |$, the potential has a negative global minimum while for larger values of 
$|\omega_I |$ the global minimum is positive. Correspondingly, there will be an upper bound on $|\omega_I |$  in order for the bound state to exist (see section \ref{shape} for further discussions).   
Furthermore, the asymptotic form of the potential is given in 
Eq. (\ref{ap-N}) and depends on whether 
$|\omega_I |>s$ or $|\omega_I |<s$. 

A special case is when $\omega= \ell=s=0$ when the effective potential 
$U_\text{eff}(N)$ (\ref{Ueff-N}) vanishes identically. This yields to the solution $A(N)=\mathrm{constant}$. Plugging this in the Eq. (\ref{Z-A}), the original field is obtained to be $Z(r(N))= (1+ a^2)^{-1} A= r$. This corresponds to a pure radial (monopole) oscillation of the BH. This solution can be obtained from the original equation (\ref{eq-pert}) as well. Plugging 
$\omega= \ell=s=0$ in   $V_{\mathrm{eff}}$ in Eq. (\ref{Veff_spin}) yields the solution $Z=c_1 r+ c_2 \ln(1- 1/r)$. However, requiring that the solution to be regular at $r=0, 1$, the only allowed solution is $Z=r$ as obtained above.
Having said this, the monopole solution is not much of physical interest.


Finally, in terms of the angular representation $a= \tan(\chi)$, and defining the new field similar to Eq. (\ref{Z-A}),  we obtain 
\ba
\label{RW-chi}
\frac{\dd^2 A(\chi)}{\dd \chi^2} - {U_\text{eff}}(\chi) A(\chi)=0 \, ,
\ea  
with  the effective potential $U_\text{eff}(\chi)$ given by,
\ba
\label{Ueff-chi}
U_\text{eff}(\chi) =  -\Big[ \frac{4 \omega^2 \cos(\chi)^6}{\sin(\chi)^2} +  (2\ell+1)^2  + \frac{1- 16 s^2 \sin(\chi)^2}{ 4 \sin(\chi)^2 \cos(\chi)^2}  
 \Big] \, .
\ea
One nice feature of the above potential is that the combination $(2 \ell +1)^2$ is separated from the rest of the potential. This is due to conformal structure of the metric (\ref{BH-metric-chi}). Furthermore, the potential 
is symmetric around the point $\chi=\frac{\pi}{2}$ if one analytically continues  the $\chi$ coordinate to the entire region $0< \chi< \pi$. 
Looking at the asymptotic regions, near the  event horizon $\chi \rightarrow 0$ and near the singularity 
$\chi \rightarrow \frac{\pi}{2}$, the potential blows up at both ends, 
\ba
\label{U-chi-ap}
U_\text{eff}(\chi \rightarrow 0) \simeq 
-\big( 4 \omega^2 +\frac{1}{4} \big) \chi^{-2} \, \quad \quad
U_\text{eff}(\chi \rightarrow \frac{\pi}{2}) \simeq \big( 4 s^2 -\frac{1}{4} \big) \big(\chi - \frac{\pi}{2} \big)^{-2} \, .
\ea

Each potential listed above has its own advantages. We shall mainly use the effective potentials $U_\text{eff}(N)$ and  $U_\text{eff}(\chi)$ given in 
Eqs. (\ref{Ueff-N}) and (\ref{Ueff-chi}).

\subsection{Bound State Perturbations}
\label{bound state}

The bound states are defined as the perturbations with $\omega^2 <0$
which are regular at the center of BH while falling off  exponentially on the event horizon so they are inaccessible to the exterior observers. To quantify this more specifically, let us look at the potential $U_{\mathrm{eff}}(N)$ in Eq. (\ref{Ueff-N}) with the asymptotic form for $N \rightarrow -\infty$ given in Eq. (\ref{ap-N}). Near the horizon $U_{\mathrm{eff}}(N) \rightarrow -4 \omega^2$ so the bound state 
solution near the horizon is given by $A(N) \sim e^{2 |\omega| N}$. 
Therefore, if $ \omega_I\equiv \mathrm{Im} (\omega) >0 $, then $A(N)$ scales like  $e^{-2i \omega N} \sim e^{-i \omega  \tau} $ and correspondingly  the total wave functions scales like  $Z(x, \tau)  \sim e^{-i \omega (x+ \tau)} $. Noting that $x+ \tau= t+ r_*\equiv v$ with $v$ being the ingoing null coordinate, we conclude that the bound state with $\omega_I>0$ represents the  ingoing perturbations  in the interior of BH \cite{Steinhauer:2025bbs}. 
On the other hand, if $\omega_I <0$, then $A(N) \sim e^{2 i \omega N}$ and $Z(x, \tau)  \sim e^{i \omega (\tau- x)} $. Noting that $x- \tau\equiv u$ with $u$ being the null outgoing coordinate, we conclude that the bound state with $\omega_I <0$ represents the outgoing bound state in the interior of BH. 
Without loss of generality,  in the rest of this work, we concentrate on the ingoing bound states  with $\omega_I >0$ and comment about the relation to the outgoing bound states in section  \ref{implications}. 

Our initial condition near the  event horizon $\tau \rightarrow -\infty$
is fixed by considering  the ingoing bound state solution with $\omega_I >0$ and $Z(x, \tau)  \sim e^{-i \omega (x+ \tau)}$.  The remaining condition is that the solution to be regular at the center
$r=0$.  In writing the starting RW equation, we employed the field $Z(x,\tau) \sim  e^{-i \omega x} Z(\tau)$. The auxiliary   field $Z(x,\tau) $ is the canonically normalized field in the sense that  if one writes the quadratic action for the perturbations 
 in the  coordinate $(\tau, x, \theta, \phi)$, then the  kinetic energy has the canonical form, i.e. it has the form 
$(\partial_\tau Z)^2/2$. However, the physical field $\Phi(\tau, x)$ is related to 
the canonically normalized field $Z(\tau, x)$ via $\Phi(\tau, x)= r^{s-1} Z(\tau, x) $. For example, for the scalar field perturbation associated to a test scalar field $\phi$, we have $\phi= Z/r$. On the other hand, the axial metric perturbation $h$  is related to $Z $ via \cite{Chandrasekhar:1985kt} $h= r Z$. Therefore, when imposing the regularity condition, we have to impose it on $\Phi$ (i.e. on $\phi$ or $h$ depending on the value of $s$)  and not on $Z$. 

 Let us look at the boundary condition for each case in more details. For scalar perturbations with $s=0$, from the form of effective potential in Eq. (\ref{Veff_spin}) we see that $V_\text{eff}(\tau) \simeq -1/4 \tau^2$. Correspondingly, the solutions near the center are 
$Z_1 = \sqrt{\tau}$ and $Z_2=\sqrt{\tau} \ln(\tau)$. However, as mentioned above, we need to impose the regularity on the physical field $\phi$ and not just on $Z$. Since $\phi= Z/r$, we conclude that the logarithmic solution $Z_2$
is not allowed and the only allowed solution near the center  is $Z_1= \sqrt{\tau} \sim r$.  This fixes the boundary condition at $r=0$ for scalar perturbations. 

Now consider the tensor perturbations with $s=2$. Solving Eq. (\ref{Veff_spin}) near the center, the solutions are $Z_1 = \tau^{3/2}$
and $Z_2 = \tau^{-1/2}$. However, we have to impose the regularity conditions on the component of metric perturbation $h= r Z \sim Z \tau^{1/2}$. The consistency of the Einstein field equations require that \cite{Chandrasekhar:1985kt}  $h \sim r^4 \sim \tau^2$ near the center. This in turn only allows the solution 
 $Z_1 = \tau^{3/2}\sim r^3$. This fixes the boundary condition at $r=0$ for tensor perturbations.

On the other hand, for the vector perturbations $s=1$, the most singular term for $V_\text{eff}(\tau)$  in Eq. (\ref{Veff_spin}) vanishes. Therefore, it is more convenient to work with the differential equation given in terms of the 
original coordinate $r$  (see Eq. \ref{eq1} in section 6). Solving the equation near $r=0$, the solutions are $Z_1= r J_2( 2 \sqrt{\ell(\ell+1)r})$ and $Z_2= r Y_2( 2 \sqrt{\ell(\ell+1)r})$ in which $J_2$ and $Y_2$ are the Bessel and  Neumann functions respectively. Using the small argument limits of $J_2$ and $Y_2$, we obtain $ Z_1\sim r^2$ while $Z_2=c$ with $c$ being a constant. Corresponds to the solutions $Z_1$ and $Z_2$, the component of the Maxwell field tensor $F_{0 \phi}$ is obtained to be \cite{Chandrasekhar:1985kt} $r^{3/2}$ and $r^{-1/2}$ respectively. The  solution $Z_2$ is therefore singular and the regular solution is only $Z_1\sim r^2$. This fixes the boundary condition for the vector perturbations. 

Summarizing the above discussions, the regularity condition for a field of spin $s$ at $r=0$ is given by  $Z \sim r^{1+s}$.

With the above discussions in mind, in next sections we solve for the spectrum of the bound state solutions. We assume that $\omega$ is pure imaginary with $\omega_I >0$. While the assumption of a pure imaginary is consistent in our setup with a real potential, but it would be interesting to examine other situations in which $\omega$ can have a real component as well. 

\subsection{The Shapes of Effective Potential }
\label{shape}

To investigate the shape of the potential, let us start with the cases $s=1, 2$
in which the potentials $U_{\mathrm{eff}}(N)$ or $U_{\mathrm{eff}}(\chi)$ have general forms. With the asymptotic form of the potential near the center given as  $U_{\mathrm{eff}}(N \rightarrow +\infty) = 4 s^2>0$ ,  the only way that the bound state can exist (i.e. the wave function falling off near the horizon $N \rightarrow -\infty$) is that  the effective potentials 
should have a negative global minimum. This requirement imposes conditions on $\omega_I$.  
To see this, let us consider $U_{\mathrm{eff}}(\chi)$ and look at the 
the point of global minimum $\chi_m$ determined via,
\ba
\label{min-eq}
32 \omega_I^2  y_m^5 - 48 \omega_I^2   y_m^4 + 
16 s^2 y_m^2 + (2- 32 s^2) y_m+ 16 s^2 -1 =0 \, , \quad \quad y\equiv \cos(\chi)^2 \, .
\ea
Solving this for $\omega_I$ yields,
\ba
\omega_I^2 = \frac{16 s^2 y_m^2 + (2- 32s^2) y_m + 16 s^2 -1}{16 y_m^4 (3- 2 y_m)} \equiv g(y_m) \, .
\ea
One can check that for the axial tensor perturbations with $s=2$ and vector perturbations with $s=1$, the minimum of $g(y_m)$ occurs at $y_m=1$ with 
the minimum value $g(1) = \frac{1}{16}$. Therefore, in order for the solution of  Eq. (\ref{min-eq}) to exist, we conclude that $\omega_I > \frac{1}{4}$. 


\begin{figure}[t]
\begin{center}
	\includegraphics[scale=0.33]{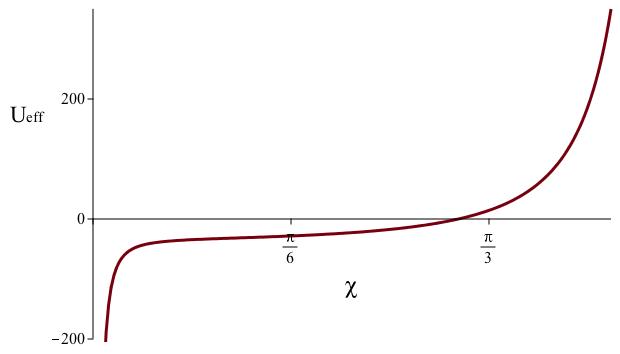}
	\hspace{1cm}
	\includegraphics[scale=0.33]{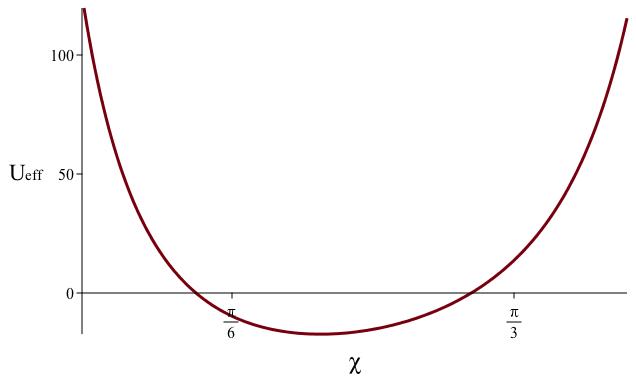}
	\end{center}
\caption{ The effective potential $U_\text{eff}(\chi)$ given in Eq. (\ref{Ueff-chi}) for $s=2$, $\ell=3$. In the left panel, $\omega_I= \frac{1}{8}$ and the bound state solution does not exit. In the right panel, $\omega_I= 1.7$ and the bound state is allowed. 
}
\label{Ueff-chi-ab}
\end{figure}

As we saw earlier, 
this conclusion is also supported from the asymptotic forms of 
$U_{\mathrm{eff}}(a)$ given in Eq. (\ref{ap-a1}) or from 
$U_{\mathrm{eff}}(\chi)$ given in  Eq. (\ref{U-chi-ap}). 
To see this more explicitly, in Fig. \ref{Ueff-chi-ab} we have plotted the effective potential $U_{\mathrm{eff}}(\chi)$ for axial perturbations with $\ell=3$ for two different values of $\omega_I$. In the left panel, $\omega_I= \frac{1}{8}$ and the effective potential diverges at the two end points with opposite signs so the bound state is not supported. On the other hand, in the right panel, $\omega_I= 1.7 $ such that the potential is negative in the intermediate region, allowing the existence of the bound state solution.

Curiously, the existence of the lower bound $\omega_I > \frac{1}{4}$ can be seen only in $U_{\mathrm{eff}}(a)$ and $U_{\mathrm{eff}}(\chi)$ but it can not be seen from the potential $U_{\mathrm{eff}}(N)$. For example, as 
the plots in Fig. \ref{Ueff-N-plot} may suggest, for  $\omega_I < \frac{1}{4}$ the global minimum is negative and  a priori there is no restriction on the existence of the bound state for  $\omega_I < \frac{1}{4}$. While this may look contradictory, we comment that there is no contradiction in the physical results. Having a negative global minimum is necessary for the existence of the bound state  but not sufficient. As we shall see in next sections, for a given value of $\ell$ (and hence a given shape of the effective potential)  the bound state exits only for limited values of $\omega_I$. 
So the fact that the lower bound $\omega_I > \frac{1}{4}$
can be deduced from $U_{\mathrm{eff}}(a)$ and $U_{\mathrm{eff}}(\chi)$ but not from $U_{\mathrm{eff}}(N)$ is not a contradiction. It just demonstrates the advantage/disadvantage of the various potentials considered here for 
the purpose of interests.

Another interesting conclusion is that for a given value of $\ell$, there is an upper bound for $\omega_I$ in order for the bound state to exist.
This is already observed in the right panel of Fig. \ref{Ueff-N-plot} in which the global minimum is uplifted to positive values when one increases  $\omega_I$. For $s=1, 2$ the existence of the upper bound is  easy to see  in terms of $U_{\mathrm{eff}}(\chi)$   in which $U_{\mathrm{eff}}(\chi)$ approaches $+\infty$ at both singular points  $\chi=0, \frac{\pi}{2}$. Therefore, there is an upper bound on $\omega_I $ in order for the potential to be  negative in the intermediate region $0< \chi < \frac{\pi}{2}$. The larger is $\ell$, the higher is the upper bound on $\omega_I$. As an estimate of the the upper bound on $\omega_I$, consider the axial perturbations ($s=2$) and  suppose $\omega_I \gg1$ so we can solve the position of the minimum from Eq. (\ref{min-eq}) approximately 
 as follows,
 \ba
 y_m \simeq \Big(  \frac{63}{48 \omega_I^2}   \Big)^{\frac{1}{4}} \, , \quad \quad (s=2 ) \, .
 \ea
Plugging the above value of the position of the minimum  into $U_{\mathrm{eff}}(\chi)$ and requiring the global minimum to be negative yields the following upper bound,
\ba
\label{upper-bound}
\omega_I^{\mathrm{max}} \lesssim \frac{4}{(21)^{\frac{3}{2}} } ( \ell + \frac{1}{2})^4  
\simeq 0.042 \,  \ell^4 \, , 
\quad \quad
\quad (s=2 ) \, .
\ea
Our numerical analysis shows that the above upper bound on $\omega_I$ works reasonably well for $\ell \gg 1$.

\begin{figure}[t]
\begin{center}
	\includegraphics[scale=0.2]{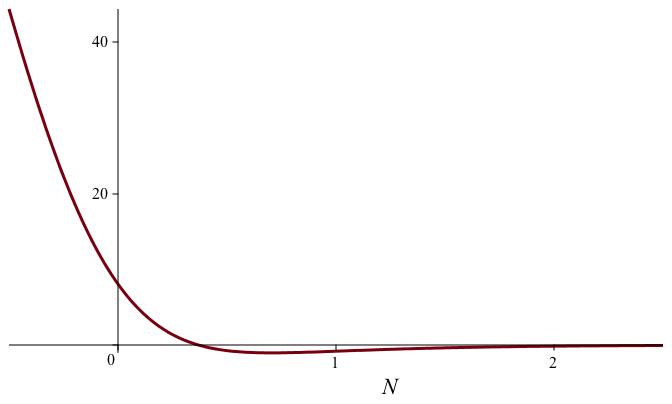}
	\includegraphics[scale=0.2]{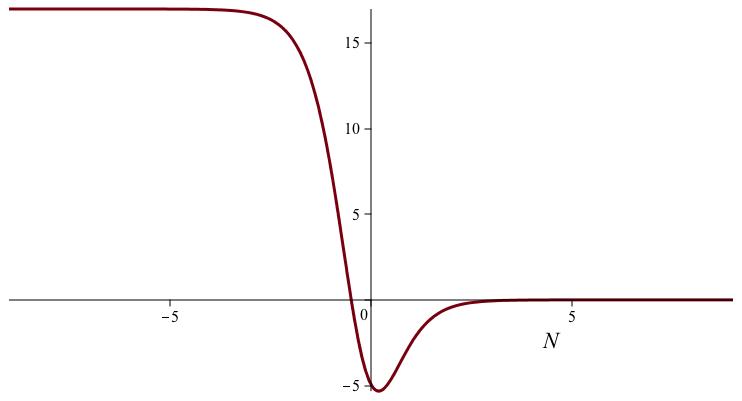}
	\includegraphics[scale=0.2]{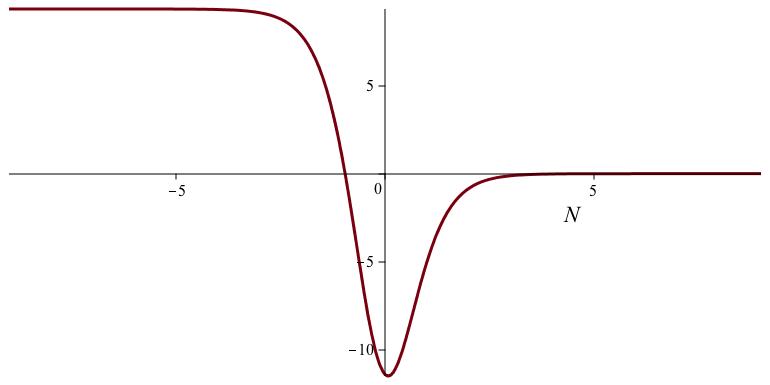}
	\end{center}
\caption{ The effective potential $U_\text{eff}(N)$ given in Eq. (\ref{Ueff-N}) for $s=0$ and various values of $\ell$.  Left: $\ell=1, \omega_I= 6.34$. Middle:  $\ell=2, \omega_I= 2.06$. Right: $\ell=3, \omega_I= 1.53$. These values of $\omega_I$ correspond to the most excited state, the lowest value of $\omega_I$ for a given $\ell$,  as given in  
table 2. 
\label{s=0-bound}
}
\end{figure}

The existence of the bound state for $s=0$ is somewhat subtle. Indeed, looking  at the effective potential $V_{\mathrm{eff}}(\tau)$ given in the original $\tau$ coordinate as shown in the left panel of Fig. \ref{Vef},   
$V_{\mathrm{eff}}(\tau)$ is always negative and  diverges at the center
of BH like $-1/4 \tau^2$.  So one may wrongly conclude that the bound state does not exist for $s=0$. However, in principle, there is no restriction for the existence of bound state for $s=0$.  Qualitatively, the situation here is  similar to the s-wave bound states for the Hydrogen atom in which the potential is negative.   Having said this, the existence of bound states for $s=0$ can be easily seen using $U_{\mathrm{eff}}(N)$ and $U_{\mathrm{eff}}(a)$
in which the potential develops a global negative minimum and 
approaches zero towards the center $a\rightarrow 0, N\rightarrow +\infty$. 
To see this property, in Fig. \ref{s=0-bound} we have plotted $U_{\mathrm{eff}}(N)$ for various values of $\ell$ with the values of $\omega_I$  obtained from the numerical analysis presented in table 2. In section \ref{numerical}, we confirm the existence of the bound state for  $s=0$ from the full numerical analysis.


\section{WKB Method}
\label{WKB}

As mentioned before, the exact solutions of the RW equation
 in terms of elementary functions are not known, so in our analytic methods, we employ various approximation schemes. In this section we employ the WKB method to find the approximate values of the  bound state spectrum.  

Consider a configuration such as in the right panel of Fig. \ref{Ueff-chi-ab} where the equation $U_{\text{eff}}(\chi)=0$ has roots at two positions 
$\chi_1$ and $\chi_2$. Noting  that in our corresponding Schrodinger-like equation the effective energy is zero, from the WKB method we obtain,
\ba
\label{WKB1}
\int_{\chi_1}^{\chi_2} \sqrt{- U_{\text{eff}}(\chi)} = 
\big( n+ \frac{1}{2} \big) \pi \, ,
\ea
with $n=0, 1,...$. 
The above integral can not be calculated analytically. However, we can approximate the potential in the region $\chi_1 \leq \chi \leq \chi_2$ to quadratic order around the global minimum $\chi_m$ as follows,
\ba
\label{WKB-ap}
U_{\text{eff}}(\chi) \simeq  U_{\text{eff}}(\chi_m) + \frac{1}{2} U_{\text{eff}}''(\chi_m)  \big( \chi- \chi_m\big)^2  \, .
\ea
We expect that the above approximation works well for the situation where the global minimum is deeply negative. This is the case for the 
ground state $n=0$ with the most negative value of $\omega^2$ (the 
largest value of $\omega_I$)  and the first excited state $n=1$ and other low-lying states. Furthermore, the larger is $\ell$, the better is  WKB's accuracy.

Using the above approximation in the WKB integral (\ref{WKB1}) and after a bit of algebra, we obtain 
\ba
\label{WKB-eq}
\frac{- U_{\text{eff}}(\chi_m)}{\sqrt{2 U_{\text{eff}}''(\chi_m)}} = 
\big( n+ \frac{1}{2} \big) \, , \quad \quad n=0, 1, 2, ... \, .
\ea
Interestingly, the above equation has the same form as the result obtained in \cite{Schutz:1985km} from the WKB method for the exterior region. 

Let us concentrate to the axial tensor perturbations for simplicity. In the limit of $\omega_I \gg1$, we have 
\ba
U_{\text{eff}}(\chi_m) \simeq 2\times (21)^\frac{3}{4} \sqrt{\omega_I} - 
4 \big(\ell + \frac{1}{2} \big)^2 \, , \quad \quad
U_{\text{eff}}''(\chi_m) \simeq 48\times \sqrt{21}\,  \omega_I \, .
 \ea
Plugging the above results in Eq. (\ref{WKB-eq}) yields, 
\ba
\label{omega-WKB}
\omega_I \simeq \frac{ 4 (  \ell +\frac{1}{2})^4}{ (21)^{\frac{3}{2}}}
\Big[ 1+ \sqrt{\frac{24}{21}} \big( n+ \frac{1}{2} \big) \Big]^{-2} \equiv f(\ell, n) \, ,  
\quad \quad   (s=2) \, .
\ea
Note that the above result is consistent with the upper bound for $\omega_I$ obtained in Eq. (\ref{upper-bound}). We see that the spectrum of $\omega_I$ scales like $(  \ell +\frac{1}{2})^4$. Furthermore, the largest value of $\omega_I$ and correspondingly, the most negative value for $\omega^2$,  
corresponds to the lowest level, $n=0$. The second  
largest value of $\omega_I$ is for $n=1$ and so on.

\begin{figure}[t]
\begin{center}
	\includegraphics[scale=0.55]{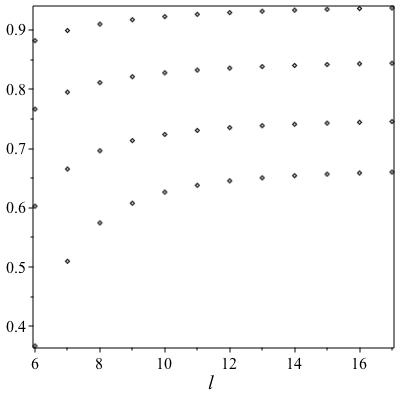}
	\end{center}
\caption{ The WKB predictions. The vertical axis denotes the ratio $\omega_I/f(\ell, n)$ with $f(\ell, n)$ obtained  from the WKB method in Eq. (\ref{omega-WKB}) while  $\omega_I$ is obtained from the full numerical analysis.  From the top row to  bottom:  $n=0, 1, 2$ and $n=3$ respectively. 
\label{WKB-plot}
}
\end{figure}

In Fig. \ref{WKB-plot} we have presented the ratio $\omega_I/f(\ell, n)$ for various value of $\ell$ for axial perturbations for the first four levels, 
$n=0, 1, 2, 3$.  We have chosen large enough value of $\ell$ to have better 
WKB accuracies.   The values of $\omega_I$ are obtained from the full numerical analysis in section \ref{numerical}.  The top to bottom rows correspond  to  $n=0, 1, 2$, and $n=3$ respectively.    As expected, the WKB result is more accurate for the lowest level $n=0$ yielding to largest value of $\omega_I$. As $n$ increases, the errors in WKB predictions grow.


\section{Method of Matching Condition}
\label{matching}

In this section we use a different approximation scheme to calculate the spectrum. We keep using the effective potential $U_{\text{eff}}(\chi)$. 
 
The method is as follows. We can not solve the differential equation with the exact potential   (\ref{Ueff-chi}). However, we expand the potential at its two end points at $\chi=0$ and $\chi=\frac{\pi}{2}$ to some orders such that reasonably good approximations of the exact solutions near each regions are obtained. Let us denote these approximate solutions by $A_-$ and $A_+$ in which $A_- (A_+)$ is the solution obtained from the expansion of the potential at $\chi=0\, (\frac{\pi}{2})$. Now, to obtain the spectrum of the perturbations, we  demand that the two solutions have overlaps in the intermediate region, $\chi = \chi_c$. We require both the amplitude and the oscillatory phases of the two solutions to match at $\chi_c$. This will fix the spectrum of  $\omega$. Of course, this approximation can be reliable only when the two solutions  have some overlaps in the intermediate  region. This requirement also  put constraints on the model parameters such as $\ell$ and $\omega$.

With the above discussions in mind, let us look at the expansions of the potential $U_{\text{eff}}(\chi)$ at its two end points.  While in Eq. (\ref{U-chi-ap}) we have expended it to the first order, here we expand it to the second order for better accuracies. At $\chi=0$, we obtain
\ba
\label{U-app1}
U_{\text{eff}}(\chi) \simeq -\big( 4 \omega^2 +\frac{1}{4} \big) \chi^{-2}
+ \frac{32 \omega^2}{3} - (2 \ell+1)^2 -\frac{1}{3} + 4 s^2 + {\cal O} (\chi^2) \, .
\ea
Solving the corresponding RW equation (\ref{RW-chi}) with the above approximation of the potential we obtain,
\ba
\label{sol-1}
A_- = C_1 \sqrt{\chi} \, J_{2  \omega_I} (\nu_- \chi)+  
C_2 \sqrt{\chi}\,  Y_{2 \omega_I} (\nu_- \chi) \, ,
\ea
in which $J_{2  \omega_I}(x)$ and  $Y_{2  \omega_I}(x)$ are the Bessel function  and the Neumann function respectively, $C_1$ and $C_2$ are two constants of integrations  while the index $\nu_-$ is defined via,
\ba
\label{nu-}
\nu_- \equiv \frac{2}{3} \sqrt{9 L + 24 \omega_I^2 - 9 s^2 + 3} \,  , 
\quad \quad  L \equiv \ell (\ell+1) \, .
\ea

On the other hand, expanding the potential at $\chi=\frac{\pi}{2}$ to second order we obtain,
\ba
\label{U-app2}
U_{\text{eff}}(\chi) \simeq \big( 4 s^2 -\frac{1}{4} \big) \big( \chi- \frac{\pi}{2}\big)^{-2}- (2 \ell+1)^2 -\frac{1}{3} + \frac{4 s^2}{3} + {\cal O} \big((\chi-\frac{\pi}{2})^2 \big) \, .
\ea
One interesting aspect of this expansion is that the spectrum $\omega_I$ does not appear to few leading orders in the expansion around $\chi=\frac{\pi}{2}$. This approximation of the potential near $\chi=\frac{\pi}{2}$ yields the approximate solution,
\ba
\label{sol-2}
A_+ =   D_1 \sqrt{\chi-\frac{\pi}{2}} J_{2 s} \Big(\nu_+ (\chi-\frac{\pi}{2}) \Big)+  D_2 \sqrt{\chi-\frac{\pi}{2}}  Y_{2 s} \Big(\nu_+ (\chi-\frac{\pi}{2}) \Big) \, ,
\ea
in which $D_1$ and $D_2$ are two new constants of integration while 
the index $\nu_+$ is given by,
\ba
\label{nu+}
\nu_+ \equiv \frac{2}{3} \sqrt{9 L -  3 s^2 + 3} \, .
\ea

The solution $A_-$ is valid near the region $\chi=0$ while $A_+$ is valid near 
$\chi=\frac{\pi}{2}$. Our goal is to extend the two solutions to intermediate region and demand that the two solutions match smoothly at an intermediate point $\chi=\chi_c$. For this purpose, we need to fix  the coefficients 
$C_i$ and $D_i$. At $\chi=0$, the solution is expected to be in the form 
$A_- \sim \chi^{2 \omega_I}$. Using the small argument limits of the Bessel and Neumann functions, it is seen that only the solution involving 
$J_{2  \omega_I} (\nu_- \chi)$ matches this asymptotic form. Therefore, we have to set $C_2=0$. Now without loss of generality we simply set $C_1=1$
and obtain,
\ba
\label{sol-1b}
A_-= \sqrt{\chi} \,  J_{2 \omega_I} (\nu_- \chi) \, .
\ea  

On the other hand, near the center of BH at $\chi=\frac{\pi}{2}$ we demand that  the solution to be regular. Looking at the small argument limit of the two functions in Eq. (\ref{sol-2}) we see that only the Bessel function $J_{2s}$ is allowed so we have to set $D_2=0$. Correspondingly,  the solution near the center is given by,
\ba
\label{sol-2b}
A_+ =   D_1 \sqrt{\chi-\frac{\pi}{2}} J_{2 s} \Big(\nu_+ (\chi-\frac{\pi}{2}) \Big).
\ea

Our job now is to propagate the two solutions (\ref{sol-1b}) and (\ref{sol-2b}) from each side towards the middle region and hope that they can smoothly match at an intermediate point. Imposing the continuity of $A_\pm$ and their derivatives $A'_\pm$ at an intermediate point $\chi_c$ yields the following constraint, 
\ba
\label{cond1}
\frac{1}{2 \chi_c} +  \frac{J_{2 \omega_I}'( \nu_- \chi_c)}{J_{2  \omega_I} ( \nu_- \chi_c  )} \, \nu_-  = 
 \frac{1}{2 (\chi_c- \frac{\pi}{2} )}  +   \frac{J_{2s}'  \big(\nu_+ (\chi_c- \frac{\pi}{2} ) \big)}{J_{2s}  \big(\nu_+ (\chi_c- \frac{\pi}{2} ) \big)}\,  \nu_+\, .
\ea
The last step is to fix the point $\chi_c$ where the two solutions are expected to overlap. This is obtained by equating the two asymptotic expansions of the potential given in Eqs. (\ref{U-app1}) and (\ref{U-app2}), yielding to the following equation for $\chi_c$, 
\ba
\label{chi-c}
( 4 \omega_I^2 -\frac{1}{4} ) \chi_c^{-2}
- \frac{32 \omega_I^2}{3}   + 4 s^2 
= \big( 4 s^2 -\frac{1}{4} \big) \big( \chi_c- \frac{\pi}{2}\big)^{-2} 
 + \frac{4 s^2}{3} \, .
\ea
Having obtained $\chi_c$ from the above equation and plugging it into Eq. (\ref{cond1}) yields  an equation which should be solved to determine $\omega_I$. 

It turns out that for a given value of $\ell$,  the current matching condition approach works well only for the lowest value of  $\omega_I$, i.e. the most excited state.  This is because the higher order terms which are neglected in Eqs. (\ref{U-app1}) become important. Furthermore, the larger is $\ell$, the better is the results from the matching condition approach. The reason is that as $\ell$ increases, the approximation  in (\ref{U-app2}) covers further region to the left  of $\chi=\frac{\pi}{2}$ 
towards $\chi=0$ and the two solutions $A_\pm$ can be overlapped more easily.  

\begin{figure}[t]
\begin{center}
	\includegraphics[scale=0.35]{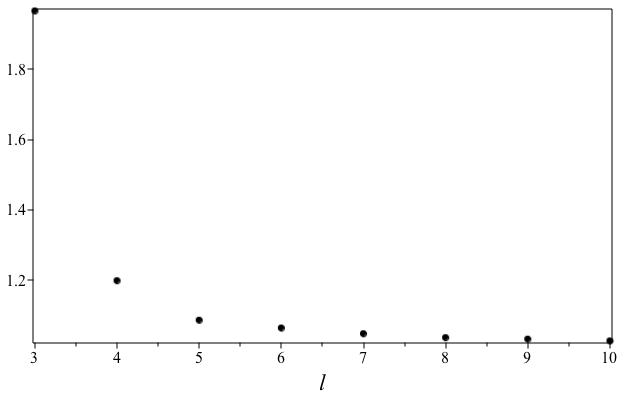}
	\hspace{0.8cm}
	\includegraphics[scale=0.35]{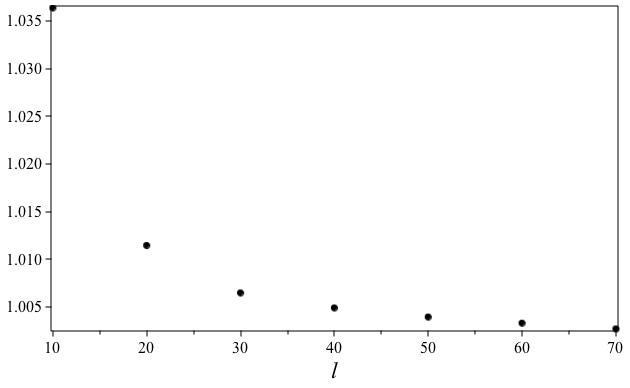}
	\end{center}
\caption{ The most excited state  (lowest value of $\omega_I$) predicted from matching condition for $s=2$.  Left: the ratio of our theoretical result for $\omega_I$ to the value of $\omega_I$ from the full numerical result. The agreement between them improves as $\ell$ increase. Right: The ratio 
$\big(  1+ \frac{1}{\ell}\big)/\omega_I$ with $\omega_I$  from the full 
numerical analysis. 
\label{matching-plot}
}
\end{figure}

In Fig. \ref{matching-plot} we have presented the predictions of the matching condition method for the most excited states associated to the axial perturbations for various values of $\ell$.  The left panel represents the ratio of  $\omega_I$ obtained from our matching condition method to the value of $\omega_I$ obtained from the full numerical analysis.  We see that as $\ell$ increases, the agreement improves. 

From our analytical method based on matching condition method, as well as from our numerical results in section \ref{numerical}, it can be seen  that for $\ell \gg 1$, 
the most excited  states approach the lower bound $\omega_I \rightarrow 1$ with the corrections at the order $\ell^{-1}$. Therefore, we conjecture the following asymptotic lower bound,
\ba
\label{conjecture}
\omega_I \rightarrow 1+ \frac{c}{\ell} + {\cal O} \big(\ell^{-2} \big) \, , \quad \quad (\ell \gg 1).
\ea
Our numerical analysis suggests that $c\simeq1$. In the right panel of Fig. \ref{matching-plot} we have plotted the ratio $\big(  1+ \frac{1}{\ell}\big)/\omega_I$ in which $\omega_I$ is the value from the full numerical analysis.
As this plot suggests, this ratio approaches unity for high values of $\ell$. These plots also demonstrate the reliability of our theoretical approach for the most excited states.


The conclusion  that $\omega_I \rightarrow 1$ for $\ell \gg 1$ can be shown explicitly as follows. In the limit $\ell \gg 1$ we have $\nu_- \simeq \nu_+ \simeq \ell $ while the Bessel function $J_\nu(x)$ can be approximated as follows,
\ba
J_\nu(x) \simeq \sqrt{\frac{2}{\pi x}} \cos\Big( x- \frac{\nu \pi}{2} - \frac{\pi}{4}
\Big) \, .
\ea
Correspondingly, the matching conditions from the continuity of the mode functions and their derivatives at the intermediate point $\chi_c =\frac{\pi}{4}$
yield,
\ba
\tan\Big( \ell \frac{\pi}{2} - \omega_I \pi -\frac{\pi}{4} \Big) 
\simeq\tan\Big( -\ell \frac{\pi}{2} - s \pi -\frac{\pi}{4} \Big)  \, .
\ea
Noting that $s=0, 1, 2$, the above equation can be satisfied only if,
\ba
\omega_I = \ell-k \, , \quad k= \ell-1, \ell-2,... \, .
\ea
The above equation suggests that with $k=\ell-1$, 
the most excited state, corresponding to the least positive value of $\omega_I$, has the value  $\omega_I= 1$ as was to be shown. Note that the above result is obtained in the limit where the subleading corrections  are discarded in the expansion of $\nu_\pm$.

The above derivation also suggests that the next highest excited states will have $\omega_I\simeq 2, 3 $ and so on. Indeed, we have checked numerically that this expectation does hold for $\ell \gg 1$. But as one proceeds with 
lower values of $k$, corresponding to larger values of $\omega_I$ (and  lower excited states),  the errors increase rapidly  and the deviations of $\omega_I$ from integer values become noticeable. For example, for axial perturbations with  $\ell=100$, the first four excited states respectively are 
$\omega_I \simeq 1.009506937, 2.038760768, 3.088919100, 4.161221778$. 
As mentioned, the errors in $\omega_I$ to be an integer value become noticeable as one moves further away from the most excited state.  

\section{Numerical Methods}
\label{numerical}

Having presented our analytical approximations in previous sections, here we present the numerical methods to obtain the spectrum of the bound state perturbations. 

Rewriting the RW equation (\ref{eq-pert}) in $r$ coordinate, we have,
\ba
\label{eq1}
r (r-1) \frac{\dd^2 Z}{\dd r^2} +  \frac{\dd Z}{\dd r} + 
\Big( \frac{\omega^2 r^3}{r-1} - L + \frac{s^2-1}{r} \Big) Z=0 \, , \quad \quad 
L\equiv \ell (\ell+1) \, .
\ea
Factoring out  the asymptotic form of the solution near the center $r=0$ and near the horizon $r=1$ via,
\ba
\label{Z-g}
Z \equiv g(r) (r-1)^{-i \omega} r^{1+s} e^{-i \omega r}\, ,
\ea
equation (\ref{eq1}), in terms of the new function $g(r)$,  is cast into,
\ba
\label{eq2}
r (r-1) \frac{\dd^2 g}{\dd r^2} +   
\Big[ -2 i \omega r^2 + 2 (s+1) r -1-2 s  \Big] \frac{\dd g}{\dd r} +
\Big[ s-L+ s^2 - 2 i (s+1) \omega r \Big] g=0 \, .
\ea
One can check that the function $g(r)$ is regular at both end points $r=0, 1$ so we employ the Frobenius series expansion as follows,
\ba
g(r)= \sum_{m=0}^{\infty} c_m r^m \, .
\ea
Plugging the above series expansion in Eq. (\ref{eq2}) yields the following
three-term  recursion relation,
\ba
\label{recurrence}
c_m =\frac{1}{m (m+2 s)} \Big[ \Big( m^2 + ( 2s-1)m + s^2 - L + s\Big) c_{m-1} - 2 i \omega (m+ s-1) c_{m-2}  \Big] \, ,
\ea
with the additional relation between $c_0$ and $c_1$,
\ba
\label{c1-0}
c_1 = \frac{s^2 + s-L}{2s+ 1} c_0 \, .
\ea

For each value of $m \ge 2$ the recursion relation (\ref{recurrence}) and the condition (\ref{c1-0}) provide a polynomial expression for $c_m$ in terms of $\omega_I$  with a normalization factor proportional to $c_0$. The above three-term recursion relation may be compared with the corresponding equation for the QNMs perturbations for the exterior of the BH 
\cite{Leaver:1985ax, Nollert:1993zz}.

By construction, $g(r)$ is regular at $r=0$ and,  in order for the function 
$g(r)$ to be regular at $r=1$,  we require that the sum 
$\sum_0^\infty c_m$ to exist and to be finite. This criteria fixes the spectrum of our bound state perturbations similar to the way that  the spectrum of QNMs in the exterior region are obtained \cite{Leaver:1985ax, Nollert:1993zz}. But the important difference is that we look for the bound state solution $\omega= i \omega_I$ with $\omega_I >0$.

We employ the numerical method advocated in \cite{Firouzjahi:2018drr} to calculate the bound state spectrum. As argued in \cite{Firouzjahi:2018drr}, in order for the sum  $\sum_0^\infty c_m$ to exist and to be finite, it is enough if for some  large value of $m$, the coefficient $c_m$ vanishes. If this condition is met for one large value of $m$, then the successive terms $c_{m+1}, c_{m+2},...$ are suppressed and one expects that the  sum $\sum_0^\infty c_m$ to converge rapidly. The good accuracy of this method compared to Leaver's method was confirmed in \cite{Firouzjahi:2018drr}. 
Following the above method, we have obtained the spectrum of the bound state solution for a given value of $\ell$ for $s=0, 1, 2$. In tables 1, 2  and  3 we have presented  $\omega_I$ in unit of $1/2 GM$ for some limited values of $\ell$ for $s= 2$ and $s=0$.

Here are the main results from our numerical analysis. 
First, our numerical analysis show that there is no bound state for $\ell=s$. 
Second, for a given value of $\ell$, there are total $\ell-s$ bound states.
It is interesting that, unlike the QNMs spectra, the number of bound state is finite.  For a given value of $\ell$, the ground state ($n=0$) has the maximum value of $\omega_I$ while the most excited state with $n= \ell-s-1$ has the least value of $\omega_I$.  Third, for large $\ell$, the spectra associated to the  first few lower states (say $n=0, 1$) satisfy the approximate analytic formula (\ref{omega-WKB}) for $s=2$.    Finally, for $\ell \gg 1$,  the spectrum for the most excited state approaches  $\omega_I \rightarrow 1$ (see table 3) as 
was observed  from our theoretical analysis in Eq. (\ref{conjecture}). 
  
\vspace{1cm}
    
\centerline{
\begin{tabular}{ |c| |l |c|c|c| c|r|}  \hline
 $n$ & $\ell=2$ &  $\ell=3$  & $\ell=4$ & $\ell=5$ &$\ell=6$& $\ell=7$  \\  \hline
0     & ~~  - & 1.705649856 & 5.745972088  & 13.76767450    & 27.78537685 & 50.20490178   \\  \hline
1& ~~ - &-                      &1.394461407  &  3.968693503    & 8.389356734 & 15.42616838  \\ \hline
2& ~~ - &-                      &    -                  & 1.274578721    &  3.315604823& 6.487859859 \\ \hline
3& ~~ - &-                      &   -                  &       -                  &  1.211013017& 2.982338127 \\ \hline
4&~~ - & ~~ -&~~ -& ~~-& ~~ -  & 1.171550415 \\ \hline
\end{tabular}  }
\vspace{1cm}

\noindent {{\bf Table 1.} $\omega_I$ for $s=2$ and for various values of 
$\ell$.  For a given value of $\ell$, the ground state (most excited state) has the maximum (minimum) value of $\omega_I$. For a given value of $\ell$, there are $\ell-2$ bound states with no bound state for $\ell=2$.  Values of $\omega_I$ here and in other tables are in unit of $( 2 GM)^{-1}$.  }

\vspace{1.5cm}

\centerline{
\begin{tabular}{ |c| |l |c|c| c| c|  r|}  \hline
 $n$ & $\ell=0$ &  $\ell=1$  & $\ell=2$ & $\ell=3$ & $\ell=4$ & $\ell=5$  \\  \hline
0& ~~  - & 6.344826792 & 47.08225968  &  178.6762366   & 485.7256436&   1081.024554\\  \hline
1& ~~ - &-                      & 2.059372644 &   8.366949092   & 23.02586787&  51.43019472\\ \hline
2& ~~ - &-                      &    -                  &   1.526213830  & 4.808040380& 
11.08559060 \\ \hline
3& ~~ - &-                      &   -                  &       -                  &   1.343001830&
3.714960023 \\  \hline
4& ~~ - &~~ - &~~ - &~~ - &~~ - & 1.252926544 \\  \hline
\end{tabular}  }
\vspace{1cm}

\noindent {{\bf Table 2.} $\omega_I$ for $s=0$. The rapid growth  of $\omega_I$ for $n=0$  scales like $(\ell+\frac{1}{2})^4$.  For a given value of $\ell$, there are $\ell$ bound states.  }

\vspace{1cm}

\centerline{
\begin{tabular}{  |l |c|c| c| c| c| c|  r|}  \hline
  $\ell=10$ &  $\ell=20$  & $\ell=30$ & $\ell=40$ & $\ell=50$ & $\ell=60$ & $\ell=70$  \\  \hline
 1.11019228  & 1.05050247 &  1.03279888 &  1.02429168   & 1.01929026 & 1.01599726  & 1.01366483   \\  \hline
\end{tabular}  }
\vspace{1cm}

\noindent {{\bf Table 3.} $\omega_I$ for the most excited states (least values of $\omega_I)$ for $s=2$ and $\ell \gg 1$.  As $\ell $ increases, $\omega_I$ approaches unity, as predicted in Eq. (\ref{conjecture}).    }

\vspace{1cm}



\begin{figure}[t]
\begin{center}
	\includegraphics[scale=0.27]{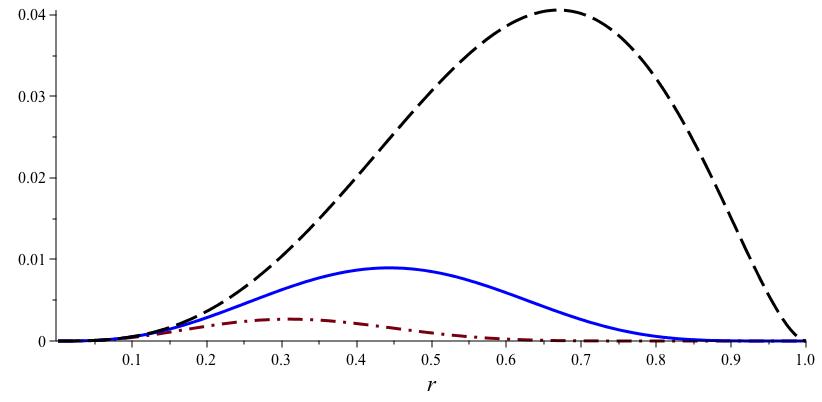}
	\hspace{1cm}
	\includegraphics[scale=0.27]{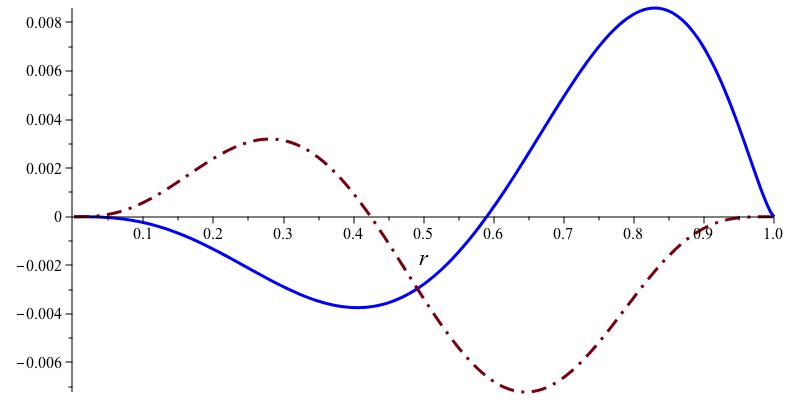}
	\end{center}
\caption{    The profile of the mode function $\Re(Z)$ for the axial perturbations. Left: the ground state ($n=0$) with no node 
for $\ell=3 $ (top, dash), $\ell=4$ (middle, solid) and $\ell=5$ (bottom,  dot-dash). Right: the  first excited state ($n=1$) with a single node, for $\ell= 4$ (solid) and $\ell=5$ (dot-dash). 
\label{profile-s=2}
}
\vspace{.5cm}
\end{figure}

\begin{figure}[t]
\begin{center}
	\includegraphics[scale=0.29]{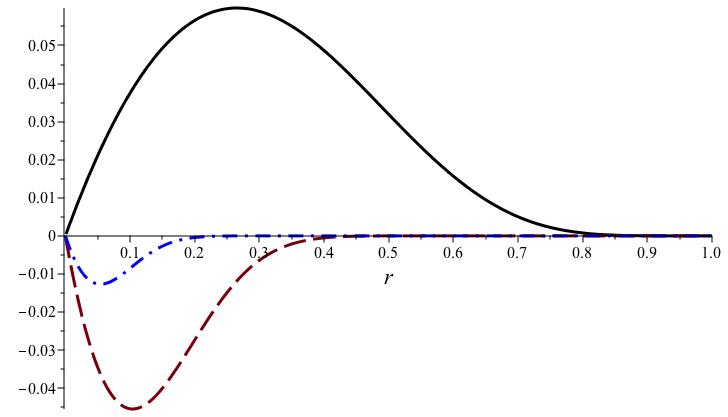}
	\hspace{1cm}
	\includegraphics[scale=0.29]{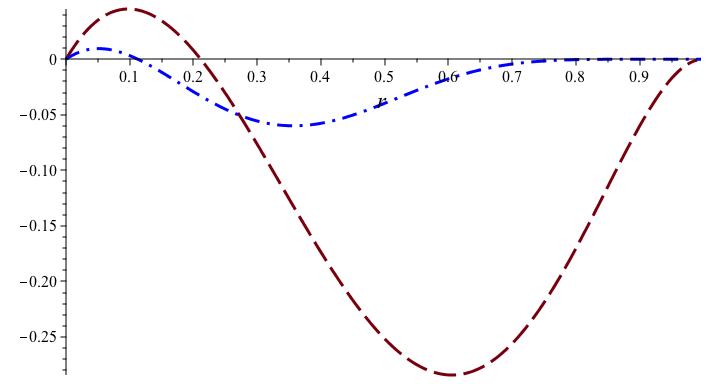}
	\end{center}
\caption{   The profile of the mode function $\Re(Z)$ for s=0. Left: the 
ground state ($n=0$) with no node for $\ell=1 $ (top, solid), $\ell=2$ (bottom, dash) and $\ell=3$ (middle,  dot-dash). Right: the first  excited state ($n=1$) with a single node, for $\ell= 2$ (dash) and $\ell=3$ (dot-dash). 
\label{profile-s=0}
\vspace{0.5cm}
}
\end{figure}

In Fig. \ref{profile-s=2} we have presented  $\Re(Z)$
for the axial perturbations in the interior of BH. In the left panel the profile of the ground state  ($n=0$) for $\ell=3, 4, 5$ are presented. As seen, the ground state has no node. In the right panel of this figure, we have presented the first excited state ($n=1$) for $\ell=4, 5$.   
As expected, the mode function for these cases has a single node. In general, the number of nodes is equal to the energy level $n$.
Correspondingly,  the most excited state with $n= \ell-3$ has $\ell-3$ nodes 
while  the next most excited state with $n=\ell-4$ has $\ell-4$ nodes and so on.

In Fig. \ref{profile-s=0} we have presented  $\Re(Z)$ for the scalar  perturbations. The left pane depicts the profile of the ground state 
$(n=0)$  for $\ell=1, 2, 3$ while the right panel shows the profile of the 
first excited state ($n=1$) for $\ell=2, 3$. The state with  level $n$ has exactly $n$ nodes.

\section{Implications}
\label{implications}

The fact that bound state can exist in the interior of BH is intriguing. These perturbations are intrinsic to the interior of BH and is not necessarily related to perturbations from the exterior regions. The existence of the bound state was originally envisaged in \cite{Firouzjahi:2018drr}. In that work, the possible links between the spectrum of the bound state perturbations and QNMs for the scalar, vector and axial tensor perturbations were studied. 
With the profile of mode function near the horizon scaling like $e^{-i \omega (x+ \tau)} = e^{\omega_I (t+ r_*)}$ it was noticed in \cite{Firouzjahi:2018drr} that the perturbations can grow on far spatial distances in the interior of BH (i.e. for $t\rightarrow \infty$). However,  it was argued that since these perturbations are not  propagating waves in the ordinary sense, the spatial growth in the form $e^{\omega_I t}$ may not cause problem. More specifically, it is argued in \cite{Firouzjahi:2018drr} that since $t$ is a space-like coordinate, one can consider a slice of constant 
$t$ and then allow the system evolves in time direction from $\tau=-\infty$ to $\tau=0$ with $\tau=-\infty$ being the near horizon regime. 

More recently, the idea of bound state in the interior of BH for axial tensor perturbations was studied in \cite{Steinhauer:2025bbs}. The interesting observation was made in \cite{Steinhauer:2025bbs}  that these bound states can have a non-zero profile on the event horizon. Because of this intriguing property,  these bound states are named overdamped quasibound states (OQBSs) in \cite{Steinhauer:2025bbs} and the  spectrum for some low-$\ell$ values are obtained. For $\ell=2$ they have reported $\omega_I \sim 10^{-5}$, and for $\ell=3, 4, 5$ they have reported $\ell-1$ bound states for each value of $\ell$ 
in which the most excited states (lowest value of $\omega_I$) have the spectrum at the order $10^{-4} $. For example, for $\ell=3$, their results are $\omega_I = 1.70573(2)$ and $\omega_I= 2.37(9)\times 10^{-4}$.
Comparing with our results for $\ell=3$, we notice that the first value of $\omega_I$ is similar to our result in table 1 (with expected deviations due to errors in numerical analysis). However, we do not find the other very small value at the order $10^{-4}$ which they have reported. Furthermore, we do not obtain the other small values of $\omega_I$ for the cases $\ell=4, 5$ either. 
Note that the values $\omega_I \ll1$ are in contradiction with our theoretical lower bound $\omega_I > \frac{1}{4}$ as observed in section \ref{perts}. 
Our analysis shows  that (as discussed in previous section) for each value of $\ell$, the number of bound states for axial perturbation is $\ell-2$, unlike the value $\ell-1$ which is  suggested in \cite{Steinhauer:2025bbs}. As we have seen, the spectrum of the bound state is always bigger than unity and it approaches unity only asymptotically for $\ell \gg1$ for the most excited state. 

In the analysis of \cite{Steinhauer:2025bbs}  it is argued that only the axial perturbations can have the OQBSs. However, our analysis shows that this property is not unique to axial perturbations and both scalar perturbations ($s=0$) and vector perturbations ($s=1$) can have bound states. 

The conclusion that the bound states can have non-zero profile at the event horizon can be understood as follows. We have solved the mode function $Z(\tau)$ assuming $\omega_I >0$ and with the boundary condition 
$e^{-i \omega r_*}$  near the future horizon. On the other hand, the total wave function has the profile $e^{-i \omega (t+ r_*)} = e^{\omega_I (t+ r_*)}$. Therefore, if one considers the far spatial distance 
on the regions near the future horizon  such that $t+r_*$ is finite, then the total mode function can acquire a non-zero value. Of course, this can happen only on the region near the corner  of the future horizon, i.e. only near the future timelike infinity $i^+$ $(U\rightarrow 0, V\rightarrow \infty)$ in the Penrose diagram as observed in  \cite{Steinhauer:2025bbs}. 

As another interesting observation, it was shown in \cite{Steinhauer:2025bbs}  that  the usual  isospectrality  between the axial and polar perturbations, which is valid for the exterior perturbations, does not hold for the OQBSs. This is because the transformations \cite{Chandrasekhar:1985kt, Chandrasekhar:1975zza, Anderson:1991kx} which link the axial and polar perturbations are singular at $r=0$. We would like to examine the spectrum of polar perturbations in a future work.

As discussed in section \ref{bound state}, for a given value of $\omega^2<0$, we have two options $\omega_I = \pm |\omega |. $ 
In this work we have considered the ingoing bound states with $\omega_I >0$ while the  outgoing bound states correspond to 
$\omega_I <0$. Motivated from the fact that the event horizon is a one-way horizon, one may conclude that the outgoing bound state is excluded. This is tricky since one may look at the outgoing solutions near the past horizon with $V=0, U>0$. Specifically, the outgoing bound state scales like 
$e^{i \omega (\tau-x)}$ so on the corner of past event horizon where $x=t\rightarrow -\infty$, one can manage a finite value of 
$u= \tau -x$ and the total profile of the outgoing bound state to be non-zero. In a senses, this suggests that  the outgoing bound state is in par with the ingoing bound state, in which the latter has a non-zero profile on the future event horizon (as observed in \cite{Steinhauer:2025bbs}) while the former has a non-zero profile on the past event horizon. Correspondingly, the above discussions imply that both ingoing and outgoing bound states can exist in the interior of BH with identical value of $|\omega_I|$.

The nature of bound state suggests that the classical energy flux from the interior of BH to the exterior of BH should vanish. More specifically, from 
Eq. (\ref{Z-g}) we note that near the horizon the mode function behaves like $Z \sim (1-r)^{\omega_I}$ while the function $g(r)$ is regular and other factors in  Eq. (\ref{Z-g}) do not play crucial roles at $r=1$. On the other hand, the components of energy density involves derivatives in the form $|\partial_r Z|^2, |\partial_x Z|^2, |\partial_\theta Z|^2$ and $|\partial_\phi Z|^2$. The first term in this list scales like $(1-r)^{2 \omega_I-1}$ while
the remaining terms scale like $(1- r)^{2 \omega_I}$. Since $\omega_I >1$ in our bound states, we conclude that at $r=1$ the classical energy flux to the  exterior of BH is zero. This is consistent with our intuition that the mode functions of the bound states are confined to the interior region and inaccessible to the exterior observers. Note that this conclusion holds at the classical level. At the quantum level, the situation can be complicated brining issues such as Hawking radiation and particle creations which are beyond the scope of our current investigation.

Before closing this section, we comment on another subtlety concerning the nature of bound states. For a given value of $\ell$, there is another set of solution for the recursion relation  Eq. (\ref{recurrence}) with all solutions having  $\omega_I <0$ (this branch of solution should not be confused with the  outgoing bound state with $\omega_I<0$ which is related to the ingoing solution by $i \omega \rightarrow -i\omega)$.  Our numerical analysis show that the number of these solutions are infinite. This set of solution corresponds to what was obtained in \cite{Firouzjahi:2018drr} in the pursuit of the bound states. However, upon close inspection of the analysis in \cite{Firouzjahi:2018drr}, it turns out that these solutions actually behave asymptotically as $e^{-|\omega_I | r_*}$ for $r_*\rightarrow -\infty$. In other words, these solutions pick up the wrong branch in the asymptotic solutions 
$e^{\pm i \omega r_*}$.  Correspondingly, 
they  blow up on the event horizon and can not be taken as the bound state solutions. Despite this sign mismatch, it is shown in \cite{Firouzjahi:2018drr} that $\omega_I$ associated to these solutions matches to good accuracy with  $\omega_I $ of the QNMs, specially for higher overtones. 
Having said this, since  $\omega_I$ has the wrong sign for these solutions to be interpreted as the bound states, the claim in \cite{Firouzjahi:2018drr} 
relating the spectrum of these solutions to  the QNMs need further justifications.

\section{Summary and Discussions}
\label{summary}

In this paper we have studied the bound state perturbations in the interior of the BH. These are defined as the perturbation with $\omega^2<0$ which are confined in the interior of BH and are inaccessible to the exterior observers. The initial condition is  that the time-dependent part of the wave function
to fall off exponentially on the event horizon while the mode function to be regular at the center of BH.   The interior of the Schwarzschild  BH is like an anisotropic cosmological background in which the azimuthal direction is shrinking while the extended spatial direction is expanding as one approaches the center of the BH.  As the relation between the conformal time $\tau$ (or the tortoise coordinate $r_*$) and $r$ is transcendental, 
we have employed the scale factor along the expanding direction as the clock and rewrote the associated RW equation in coordinates $N$ or $\chi$  with the effective potentials $U_{\mathrm{eff}}(N)$ and $U_{\mathrm{eff}}(\chi)$ respectively which have algebraic expressions. 

We have studied the spectrum of the ingoing bound states 
(with $\omega_I>0)$ for the scalar, vector and axial tensor perturbations. As the solutions for the corresponding RW equation can not be expressed in terms of standard elementary functions, we have employed analytical approximations to calculate the spectrum. This includes the WKB method and the matching condition method. The WKB method is more reliable for the lowest states (say $n=0, 1$) which have the highest values of $\omega_I$. On the other hand, the matching condition method is reliable for the most excited state (corresponding to lower value of $\omega_I$) and for large values of $\ell$. The fact that these two analytic methods cover the two opposite sides of spectrum is helpful for theoretical estimation of $\omega_I$.  

It is verified that the bound state solutions exist for 
all three types of perturbations (scalar, vector and axial perturbations). We have shown that for a given value of $ \ell$, there are total $\ell-s$ bound states. Correspondingly, for axial perturbations 
there are $\ell-2$ bound states while for scalar perturbations this is equal to 
$\ell$. The number of nodes in the profile of the bound states 
is equal to the energy level $n$. For example, the ground state with $n=0$ has no node, while the  first excited state with $n=1$ has one node and so on. We have shown both numerically and analytically that there is a universal lower bound for the spectrum of bound state  $2 G M \omega_I > 1$. This lower bound is only asymptotically saturated in the eikonal limit 
$\ell  \gg 1$. On the other hand, for a given value of $\ell$,
there is an upper bound for the spectrum of bound state which for axial perturbations and for $\ell \gg 1$ is approximately given by $2 G M \omega_I^{\mathrm{max} } \lesssim 0.042\,  \ell^4$. 
Furthermore, we have shown analytically and confirmed numerically that the spectrum of the 
ground state as well as  the nearby low-level states scale like $(\ell+ \frac{1}{2})^4$. Finally, as observed in \cite{Steinhauer:2025bbs}, these bound states have the curious property that they can have a non-zero profile on the regions near the future event horizon. 

There are a number of directions in which the current study can be extended. An interesting question is to study the bound states associated to the polar perturbations. In this case one has to look at the Zerilli equation which is more complicated than the RW equation. As the transformation relating the polar and axial perturbations is singular at $r=0$, the isospectrality condition may not hold \cite{Steinhauer:2025bbs}. Therefore, it is an open question 
whether or not our conclusions about the spectrum of axial perturbations summarized above are valid for polar perturbations as well. Another question of interest is the quantum aspects of the bound state perturbations. The analysis performed here were purely at the classical level. However, it is an interesting question to  impose the quantization conditions for the perturbations in the interior of BH and examine if there are relations between the bound state perturbations and other known quantum effects such as the Hawking radiation.

\vspace{1.5cm}

{\bf Acknowledgments:}  We thank  Bahram Mashhoon, Sebastian V{\"o}lkel, Borna Salehian and  Masoud Molaei   for useful discussions and correspondences. We thank Kazem Rezazadeh for the assistance in numerical analysis. We acknowledged using Maple 2017 software to perform the numerical analysis. 
This work is supported by the INSF  of Iran under the grant  number 4046375.



\bibliography{BH-interior}{}

\providecommand{\href}[2]{#2}\begingroup\raggedright\begin{thebibliography}{10}

\bibitem{LIGOScientific:2016aoc}
{\scshape LIGO Scientific, Virgo} collaboration, B.~P. Abbott et~al.,
  \emph{{Observation of Gravitational Waves from a Binary Black Hole Merger}},
  \href{https://doi.org/10.1103/PhysRevLett.116.061102}{\emph{Phys. Rev. Lett.}
  {\bfseries 116} (2016) 061102},
  [\href{https://arxiv.org/abs/1602.03837}{{\ttfamily 1602.03837}}].

\bibitem{LIGOScientific:2017vwq}
{\scshape LIGO Scientific, Virgo} collaboration, B.~P. Abbott et~al.,
  \emph{{GW170817: Observation of Gravitational Waves from a Binary Neutron
  Star Inspiral}},
  \href{https://doi.org/10.1103/PhysRevLett.119.161101}{\emph{Phys. Rev. Lett.}
  {\bfseries 119} (2017) 161101},
  [\href{https://arxiv.org/abs/1710.05832}{{\ttfamily 1710.05832}}].

\bibitem{LIGOScientific:2020zkf}
{\scshape LIGO Scientific, Virgo} collaboration, R.~Abbott et~al.,
  \emph{{GW190814: Gravitational Waves from the Coalescence of a 23 Solar Mass
  Black Hole with a 2.6 Solar Mass Compact Object}},
  \href{https://doi.org/10.3847/2041-8213/ab960f}{\emph{Astrophys. J. Lett.}
  {\bfseries 896} (2020) L44},
  [\href{https://arxiv.org/abs/2006.12611}{{\ttfamily 2006.12611}}].

\bibitem{LIGOScientific:2025obp}
{\scshape LIGO Scientific, VIRGO, KAGRA} collaboration, \emph{{Black Hole
  Spectroscopy and Tests of General Relativity with GW250114}},
  \href{https://arxiv.org/abs/2509.08099}{{\ttfamily 2509.08099}}.

\bibitem{Nollert:1999ji}
H.-P. Nollert, \emph{{TOPICAL REVIEW: Quasinormal modes: the characteristic
  `sound' of black holes and neutron stars}},
  \href{https://doi.org/10.1088/0264-9381/16/12/201}{\emph{Class. Quant. Grav.}
  {\bfseries 16} (1999) R159--R216}.

\bibitem{Kokkotas:1999bd}
K.~D. Kokkotas and B.~G. Schmidt, \emph{{Quasinormal modes of stars and black
  holes}}, \href{https://doi.org/10.12942/lrr-1999-2}{\emph{Living Rev. Rel.}
  {\bfseries 2} (1999) 2},
  [\href{https://arxiv.org/abs/gr-qc/9909058}{{\ttfamily gr-qc/9909058}}].

\bibitem{Berti:2009kk}
E.~Berti, V.~Cardoso and A.~O. Starinets, \emph{{Quasinormal modes of black
  holes and black branes}},
  \href{https://doi.org/10.1088/0264-9381/26/16/163001}{\emph{Class. Quant.
  Grav.} {\bfseries 26} (2009) 163001},
  [\href{https://arxiv.org/abs/0905.2975}{{\ttfamily 0905.2975}}].

\bibitem{Berti:2025hly}
E.~Berti et~al., \emph{{Black hole spectroscopy: from theory to experiment}},
  \href{https://arxiv.org/abs/2505.23895}{{\ttfamily 2505.23895}}.

\bibitem{Regge:1957td}
T.~Regge and J.~A. Wheeler, \emph{{Stability of a Schwarzschild singularity}},
  \href{https://doi.org/10.1103/PhysRev.108.1063}{\emph{Phys. Rev.} {\bfseries
  108} (1957) 1063--1069}.

\bibitem{Zerilli:1970se}
F.~J. Zerilli, \emph{{Effective potential for even parity Regge-Wheeler
  gravitational perturbation equations}},
  \href{https://doi.org/10.1103/PhysRevLett.24.737}{\emph{Phys. Rev. Lett.}
  {\bfseries 24} (1970) 737--738}.

\bibitem{Chandrasekhar:1985kt}
S.~Chandrasekhar, \emph{{The mathematical theory of black holes}}.
\newblock 1985.

\bibitem{Chandrasekhar:1975zza}
S.~Chandrasekhar and S.~L. Detweiler, \emph{{The quasi-normal modes of the
  Schwarzschild black hole}},
  \href{https://doi.org/10.1098/rspa.1975.0112}{\emph{Proc. Roy. Soc. Lond. A}
  {\bfseries 344} (1975) 441--452}.

\bibitem{Leaver:1985ax}
E.~W. Leaver, \emph{{An Analytic representation for the quasi normal modes of
  Kerr black holes}}, \href{https://doi.org/10.1098/rspa.1985.0119}{\emph{Proc.
  Roy. Soc. Lond. A} {\bfseries 402} (1985) 285--298}.

\bibitem{Anderson:1991kx}
A.~Anderson and R.~H. Price, \emph{{Intertwining of the equations of black hole
  perturbations}}, \href{https://doi.org/10.1103/PhysRevD.43.3147}{\emph{Phys.
  Rev. D} {\bfseries 43} (1991) 3147--3154}.

\bibitem{Nollert:1993zz}
H.-P. Nollert, \emph{{Quasinormal modes of Schwarzschild black holes: The
  determination of quasinormal frequencies with very large imaginary parts}},
  \href{https://doi.org/10.1103/PhysRevD.47.5253}{\emph{Phys. Rev. D}
  {\bfseries 47} (1993) 5253--5258}.

\bibitem{Mashhoon:1982im}
B.~Mashhoon, \emph{{QUASINORMAL MODES OF A BLACK HOLE}},  in \emph{{3rd Marcel
  Grossmann Meeting on the Recent Developments of General Relativity}}, 1982.

\bibitem{Blome:1981azp}
H.-J. Blome and B.~Mashhoon, \emph{{Quasi-normal oscillations of a
  schwarzschild black hole}},
  \href{https://doi.org/10.1016/0375-9601(84)90769-2}{\emph{Phys. Lett. A}
  {\bfseries 100} (1981) 231--234}.

\bibitem{Ferrari:1984zz}
V.~Ferrari and B.~Mashhoon, \emph{{New approach to the quasinormal modes of a
  black hole}}, \href{https://doi.org/10.1103/PhysRevD.30.295}{\emph{Phys. Rev.
  D} {\bfseries 30} (1984) 295--304}.

\bibitem{Ferrari:1984ozr}
V.~Ferrari and B.~Mashhoon, \emph{{Oscillations of a Black Hole}},
  \href{https://doi.org/10.1103/PhysRevLett.52.1361}{\emph{Phys. Rev. Lett.}
  {\bfseries 52} (1984) 1361}.

\bibitem{Liu:1996cxr}
H.~Liu and B.~Mashhoon, \emph{{On the spectrum of oscillations of a
  Schwarzschild black hole}},
  \href{https://doi.org/10.1088/0264-9381/13/2/012}{\emph{Class. Quant. Grav.}
  {\bfseries 13} (1996) 233--251}.

\bibitem{Schutz:1985km}
B.~F. Schutz and C.~M. Will, \emph{{BLACK HOLE NORMAL MODES: A SEMIANALYTIC
  APPROACH}}, \href{https://doi.org/10.1086/184453}{\emph{Astrophys. J. Lett.}
  {\bfseries 291} (1985) L33--L36}.

\bibitem{Iyer:1986np}
S.~Iyer and C.~M. Will, \emph{{Black Hole Normal Modes: A {WKB} Approach. 1.
  Foundations and Application of a Higher Order {WKB} Analysis of Potential
  Barrier Scattering}},
  \href{https://doi.org/10.1103/PhysRevD.35.3621}{\emph{Phys. Rev. D}
  {\bfseries 35} (1987) 3621}.

\bibitem{Iyer:1986nq}
S.~Iyer, \emph{{BLACK HOLE NORMAL MODES: A WKB APPROACH. 2. SCHWARZSCHILD BLACK
  HOLES}}, \href{https://doi.org/10.1103/PhysRevD.35.3632}{\emph{Phys. Rev. D}
  {\bfseries 35} (1987) 3632}.

\bibitem{Motl:2003cd}
L.~Motl and A.~Neitzke, \emph{{Asymptotic black hole quasinormal frequencies}},
  \href{https://doi.org/10.4310/ATMP.2003.v7.n2.a4}{\emph{Adv. Theor. Math.
  Phys.} {\bfseries 7} (2003) 307--330},
  [\href{https://arxiv.org/abs/hep-th/0301173}{{\ttfamily hep-th/0301173}}].

\bibitem{Andersson:2003fh}
N.~Andersson and C.~J. Howls, \emph{{The Asymptotic quasinormal mode spectrum
  of nonrotating black holes}},
  \href{https://doi.org/10.1088/0264-9381/21/6/021}{\emph{Class. Quant. Grav.}
  {\bfseries 21} (2004) 1623--1642},
  [\href{https://arxiv.org/abs/gr-qc/0307020}{{\ttfamily gr-qc/0307020}}].

\bibitem{Berti:2005ys}
E.~Berti, V.~Cardoso and C.~M. Will, \emph{{On gravitational-wave spectroscopy
  of massive black holes with the space interferometer LISA}},
  \href{https://doi.org/10.1103/PhysRevD.73.064030}{\emph{Phys. Rev. D}
  {\bfseries 73} (2006) 064030},
  [\href{https://arxiv.org/abs/gr-qc/0512160}{{\ttfamily gr-qc/0512160}}].

\bibitem{Hatsuda:2019eoj}
Y.~Hatsuda, \emph{{Quasinormal modes of black holes and Borel summation}},
  \href{https://doi.org/10.1103/PhysRevD.101.024008}{\emph{Phys. Rev. D}
  {\bfseries 101} (2020) 024008},
  [\href{https://arxiv.org/abs/1906.07232}{{\ttfamily 1906.07232}}].

\bibitem{Volkel:2025lhe}
S.~H. V{\"o}lkel, \emph{{Bound States of the Schwarzschild Black Hole}},
  \href{https://doi.org/10.1103/tbm2-gzv9}{\emph{Phys. Rev. Lett.} {\bfseries
  134} (2025) 241401}, [\href{https://arxiv.org/abs/2505.17186}{{\ttfamily
  2505.17186}}].

\bibitem{Hod:2025scz}
S.~Hod, \emph{{Bound-state resonances of the Schwarzschild black hole: Analytic
  treatment}}, \href{https://doi.org/10.1103/l5dw-ddrw}{\emph{Phys. Rev. D}
  {\bfseries 112} (2025) 064040},
  [\href{https://arxiv.org/abs/2506.15768}{{\ttfamily 2506.15768}}].

\bibitem{Firouzjahi:2018drr}
H.~Firouzjahi, \emph{{The spectrum of perturbations inside the Schwarzschild
  black hole}},  \href{https://arxiv.org/abs/1805.11289}{{\ttfamily
  1805.11289}}.

\bibitem{Steinhauer:2025bbs}
J.~Steinhauer, K.~Destounis and R.~Brito, \emph{{Dark tranquility: Overdamped
  quasibound states inside a Schwarzschild black hole}},
  \href{https://arxiv.org/abs/2509.02676}{{\ttfamily 2509.02676}}.

\bibitem{Fiziev:2006tx}
P.~P. Fiziev, \emph{{In the exact solutions of the Regge-Wheeler equation in
  the Schwarzschild black hole interior}},
  \href{https://arxiv.org/abs/gr-qc/0603003}{{\ttfamily gr-qc/0603003}}.

\bibitem{Firouzjahi:2024xbo}
H.~Firouzjahi, \emph{{Quantum fluctuations in the interior of black holes and
  backreactions}},
  \href{https://doi.org/10.1103/PhysRevD.110.025022}{\emph{Phys. Rev. D}
  {\bfseries 110} (2024) 025022},
  [\href{https://arxiv.org/abs/2405.05750}{{\ttfamily 2405.05750}}].

\bibitem{Kodama:1984ziu}
H.~Kodama and M.~Sasaki, \emph{{Cosmological Perturbation Theory}},
  \href{https://doi.org/10.1143/PTPS.78.1}{\emph{Prog. Theor. Phys. Suppl.}
  {\bfseries 78} (1984) 1--166}.

\bibitem{Mukhanov:1990me}
V.~F. Mukhanov, H.~A. Feldman and R.~H. Brandenberger, \emph{{Theory of
  cosmological perturbations. Part 1. Classical perturbations. Part 2. Quantum
  theory of perturbations. Part 3. Extensions}},
  \href{https://doi.org/10.1016/0370-1573(92)90044-Z}{\emph{Phys. Rept.}
  {\bfseries 215} (1992) 203--333}.

\bibitem{Kantowski:1966te}
R.~Kantowski and R.~K. Sachs, \emph{{Some spatially homogeneous anisotropic
  relativistic cosmological models}},
  \href{https://doi.org/10.1063/1.1704952}{\emph{J. Math. Phys.} {\bfseries 7}
  (1966) 443}.

\bibitem{Doran:2006dq}
R.~Doran, F.~S.~N. Lobo and P.~Crawford, \emph{{Interior of a Schwarzschild
  black hole revisited}},
  \href{https://doi.org/10.1007/s10701-007-9197-6}{\emph{Found. Phys.}
  {\bfseries 38} (2008) 160--187},
  [\href{https://arxiv.org/abs/gr-qc/0609042}{{\ttfamily gr-qc/0609042}}].

\bibitem{Firouzjahi:2022rtn}
H.~Firouzjahi and A.~Talebian, \emph{{White hole cosmology and Hawking
  radiation from quantum cosmological perturbations}},
  \href{https://doi.org/10.1103/PhysRevD.106.123505}{\emph{Phys. Rev. D}
  {\bfseries 106} (2022) 123505},
  [\href{https://arxiv.org/abs/2210.15186}{{\ttfamily 2210.15186}}].

\bibitem{Gangopadhyaya:2017wpf}
A.~Gangopadhyaya, J.~V. Mallow and C.~Rasinariu, \emph{{Supersymmetric Quantum
  Mechanics}: {An Introduction}}.
\newblock World Scientific, 2017,
  \href{https://doi.org/10.1142/10475}{10.1142/10475}.

\end{thebibliography}\endgroup

\bibliographystyle{JHEP}

\end{document}